%% file: main.tex
\documentclass[reqno]{amsart}

\usepackage{amsthm,amsmath,amsfonts,amssymb,epsfig,graphicx,latexsym,float,color}
\usepackage{subcaption}
\usepackage{enumitem}
\usepackage{pgfplots}
\usepackage{physics}
\usepackage{multirow}
\usetikzlibrary{plotmarks}
\usetikzlibrary{external}

\newtheorem{definition}{Definition}

\newtheorem{remark}[definition]{Remark}
\newtheorem{assumption}[definition]{Assumption}

\newtheorem{algorithm}[definition]{Algorithm}
\newtheorem{system}{System}

\renewcommand{\theequation}{\arabic{section}.\arabic{equation}}
\renewcommand{\thetable}{\arabic{section}.\arabic{table}}
\renewcommand{\thefigure}{\arabic{section}.\arabic{figure}}

\newcommand{\tn}[1]{\textnormal{#1}}
\newcommand{\bs}[1]{\boldsymbol{#1}}

\newcommand{\deriv}[1]{\frac{\mathrm{d} #1}{\mathrm{d} s}}

\newcommand{\closureFrU}{\mathcal{U}_{r,u}}
\newcommand{\closureFzU}{\mathcal{U}_{z,u}}
\newcommand{\closureFzUTilde}{\tilde{\mathcal{U}}_{z,u}}
\newcommand{\closureFrRho}{\mathcal{U}_{r,\rho}}
\newcommand{\closureFzRho}{\mathcal{U}_{z,\rho}}
\newcommand{\closureFzRhoTilde}{\tilde{\mathcal{U}}_{z,\rho}}
\newcommand{\abkU}{\mathcal{R}_u} 
\newcommand{\abkRho}{\mathcal{R}_\rho} 

\newcommand{\vece}[1]{\vec{\boldsymbol{#1}}}

\newcommand{\tensor}[1]{\bar{\bar{\bs{#1}}}}

\newcommand{\dr}{\;\mathrm{d}r}

\newcommand{\avg}[1]{\left\langle #1 \right\rangle_\mathcal{A}}
\newcommand{\avgBound}[1]{\left\langle #1 \right\rangle_{\partial \mathcal{A}}}
\newcommand{\avgUnit}[1]{\mathfrak{I}\left(#1\right)}

\begin{document}

\title[]{Viscoelastic model hierarchy for fiber melt spinning of semi-crystalline polymers}
\author[Ettm\"uller et al.]{Manuel Ettm\"uller$^{1}$}
\author[]{Walter Arne$^{1}$}
\author[]{Nicole Marheineke$^{2}$}
\author[]{Raimund Wegener$^{1}$}

\date{\today\\
$^1$ Fraunhofer ITWM, Fraunhofer Platz 1, D-67663 Kaiserslautern, Germany\\
$^2$ Universit\"at Trier, Lehrstuhl Modellierung und Numerik, Universit\"atsring 15, D-54296 Trier, Germany}

\begin{abstract}
In the fiber melt spinning of semi-crystalline polymers, the degree of crystallization can be non-homogeneous over the cross-section of the fiber, affecting the properties of the end product. For simulation-based process design, the question arises as to which fiber quantities and hence model equations must be resolved in radial direction to capture all practically relevant effects and at the same time imply a model that can be computed with reasonable effort. In this paper, we present a hierarchy of viscoelastic two-phase fiber models ranging from a complex, fully resolved and highly expensive three-dimensional description to a cross-sectionally averaged, cheap-to-evaluate one-dimensional model. In particular, we propose a novel stress-averaged one-two-dimensional fiber model, which circumvents additional assumptions on the inlet profiles needed in the established stress-resolved fiber model by Doufas et al.\ (2001). Simulation results demonstrate the performance and application regime of the dimensionally reduced models. The novel stress-averaged variant provides fast and reliable results, especially in the regime of low flow-enhanced crystallization.
\end{abstract}

\maketitle

\noindent
{\sc Keywords.}Crystallization; fiber spinning; melt spinning; viscoelastic two-phase model; model hierarchy; boundary value problems \\
{\sc AMS-Classification.} 76-XX; 34Bxx; 34E15; 65L10; 68U20; 35Q79

\setcounter{equation}{0} \setcounter{figure}{0} \setcounter{table}{0}
\section{Introduction}\label{sec:introduction}
Fiber melt spinning is one of the most important steps in the production of technical textiles. In the case of semi-crystalline polymers, crystallization takes place along the spinline, which has a significant influence on the properties of the end product. For design and optimization of the industrial process modeling and simulation are essential in order to reduce time and material costs. The fiber behavior in the surrounding air stream can be described by a three-dimensional multiphase-multiscale model. However, despite current high-performance computers, the direct simulation is very expensive due to the model complexity and is not possible for industrial conditions. The challenge is therefore to establish models that cover as many physically relevant effects as possible and can be simulated at the same time with reasonable computing effort.

The trending fiber spinning model for semi-crystalline polymers comes from Doufas et al.~ \cite{doufas_simulation_2001,doufas_simulation_2000-1, doufas_simulation_2000}. The stationary uniaxial viscoelastic two-phase model consists in its original version of cross-sectionally averaged balances for mass, momentum and energy. The amorphous and semi-crystalline phases are described with separate constitutive equations, whereby the amorphous phase is treated as a modified Giesekus fluid and the semi-crystalline phase as a collection of rigid rods. The transition between the phases is realized by an evolution equation for the crystallinity, and the coupling with the surrounding air is incorporated via exchange source terms for aerodynamic drag force and heat transfer. Shrikhande et al.~\cite{shrikhande_modified_2006} proposed a crystallization rate that depends on the stored free energy of the melt phase. Both one-dimensional (1D) models are well-established and are successfully used to simulate Nylon and PET fibers. As shown in \cite{ettmuller_asymptotic_2021}, they can be treated within a common model class, and the derivation of asymptotically justified boundary conditions eliminated the originally occurring boundary layers at the inlet. Extensions to this model class include radial profiles of temperature, stress and crystallinity, which lead to a description with coupled one- and two-dimensional model equations, cf.~ \cite{doufas_two-dimensional_2001}. We refer to it as \emph{stress-resolved one-two-dimensional (1D2D) model}. The asymptotically justified boundary conditions of the 1D model class are straightforward applicable to it, cf.\ \cite{ettmueller_ecmi_2024}. The consideration of radial profiles is necessary, because the degree of crystallization in the fiber cross-sections may not be homogeneous, as observed in experiments \cite{pan_radial_2019}. Other approaches of radially resolved fiber models can be found in \cite{henson_thin-filament_1998,hutchenson_radial_1984,kohler_2d_2007,matsuo_studies_1976,ottone_numerical_2002,sun_numerical_2000,vassilatos_issues_1992}.

In this work we theoretically and numerically investigate an aspect in the 1D2D model extension of Doufas et al.~\cite{doufas_two-dimensional_2001}. Starting from a three-dimensional version of the viscoelastic two-phase fiber model, the derivation of the stress-resolved 1D2D model shows that in \cite{doufas_two-dimensional_2001} additional assumptions were made regarding the inlet profiles. In particular, the choice of a constant inlet profile for the components of the conformation tensor in the amorphous phase cannot be derived directly from the 3D model, but is an assumption. In this paper we propose an alternative 1D2D extension that does not require any further assumptions on the boundary conditions. By cross-sectionally averaging all model equations related to the extra stress tensor components we obtain a description to which we refer to as \emph{stress-averaged 1D2D model}. We embed it into a model hierarchy for fiber spinning, ranging from the complex three-dimensional description over the well-established stress-resolved 1D2D version to the existing cross-sectionally averaged 1D models. We verify the validity of the 1D2D model extensions and in particular check the inlet profile assumption made in \cite{doufas_two-dimensional_2001} by comparing the model outcomes with a 3D reference for a simple test scenario. In addition, we discuss the application regime and the performance (accuracy vs. computational effort) of the simulations for practice-relevant scenarios in order to open up the field for simulation-based process design.

The paper is structured as follows. In Section~\ref{sec:3D} we prescribe the three-dimensional reference model for the dynamics of a single fiber. The dimensionally reduced models of our model hierarchy are presented in Section~\ref{sec:reducedModels}. The two 1D2D models are partially cross-sectionally averaged, whereas the 1D model results from cross-sectionally averaging all model equations. In Section~\ref{sec:compareFenics} we compare the simulation outcomes of the dimensionally reduced models and the 3D reference in terms of approximation quality and computational effort. Their performance in a practically relevant Nylon test case is investigated in Section~\ref{sec:nylonCase}. The appendix provides further details on the derivation of the model hierarchy, the used closure models, the test case setup and the numerical implementation.

\section{Three-dimensional viscoelastic two-phase fiber model} \label{sec:3D}
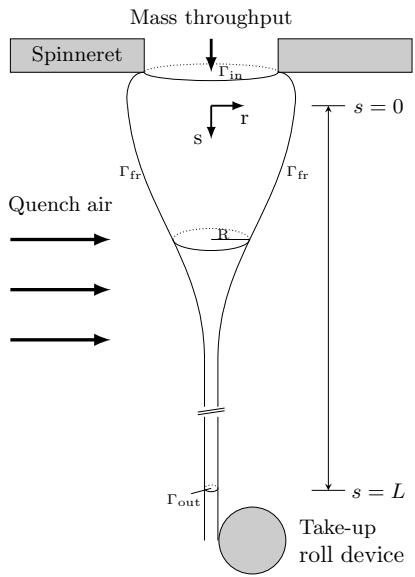
\begin{figure}[t]
\centering
\scalebox{0.89}{\input{Tikz/sketchRadialMeltSpinning.tikz}}
\caption{Fiber in melt spinning process. The upper part of the spinline near the nozzle and the radial profile are shown enlarged.}
\label{fig:sketchMeltSpinning}
\end{figure}

In the fiber melt spinning process, thin hot polymer jets are extruded through small orifices in the spinneret, cooled down by a quench air stream and drawn down by a take-up roll device. In this work we focus on the behavior of a single fiber in a stationary set-up, see Fig.~\ref{fig:sketchMeltSpinning}.

Let $\mathcal{Q} \in \mathbb{E}^3$ be the a priori unknown fiber domain in the three-dimensional Euclidian space $\mathbb{E}^3$, where the boundary $\partial \mathcal{Q} = \Gamma_\tn{in} \cup \Gamma_\tn{out} \cup \Gamma_\tn{fr}$ consists of the inlet $\Gamma_\tn{in}$, the outlet $\Gamma_\tn{out}$ and the free lateral surface $\Gamma_\tn{fr}$. The fiber flow can be modeled with two polymer phases, i.e., an amorphous phase and a semi-crystalline phase, that act in parallel with the same mass density $\rho$, velocity $\vece{v}$ and temperature $T$ and are prescribed by different material laws. The amorphous phase is treated as a Giesekus-type fluid and the semi-crystalline phase as a collection of rigid rods. Then, the two-phase fiber model combines the balances for mass, momentum and energy with constitutive equations for the conformation tensor $\tensor{c}$ in the amorphous phase and with constitutive equations for the orientational tensor $\tensor{S}$ in the semi-crystalline phase. The transition between the phases can be described by evolution equations for the crystallinity $x$ and the stored free energy $a$ in case of an energy-driven crystallization rate. 

In the spinning process the fiber is exposed to gravity and a surrounding cooling air stream. Inlet conditions are prescribed, and at the outlet the fiber velocity is given by the take-up roll device. The stationary free boundary value problem for fiber spinning is given by System~\ref{system_3d}.
\begin{system}[Three-dimensional fiber model] \label{system_3d} ~\newline
\begin{subequations} 
Balance laws in $\mathcal{Q}$:
\begin{align*}
\nabla \cdot (\rho\vece{ v}) &= 0, \\
\nabla \cdot (\rho \vece{v} \otimes \vece{v}) &= \nabla \cdot \tensor{\Sigma}^T + \vece{g}, \\
c_\mathrm{p} \rho \vece{v} \cdot \nabla T &= \nabla \cdot (C \nabla T) +  \Phi_\infty \Delta H_\mathrm{f} \rho \vece{v} \cdot \nabla x, \\
\vece{v} \cdot \nabla x &= K (1-x), \\
\vece{v} \cdot \nabla a &= - \frac{1}{\lambda_\mathrm{am}} a + \frac{G}{\zeta} \frac{\langle \tensor{c}, \nabla \vece{v}\rangle_F}{1-x}. \\
\intertext{Kinematic, dynamic and heat flux boundary conditions at the free surface $\Gamma_{\mathrm{fr}}$:}
\vece{v} \cdot \vece{n} &= 0, \\
\tensor{\Sigma} \cdot \vece{n} &= \vece{f}_\mathrm{air} + \vece{f}_\mathrm{st}, \\
-C \nabla T \cdot \vece{n} &= \alpha_\mathrm{T} (T-T_\mathrm{air}).
\end{align*}
Constitutive laws:
\begin{align*}
\tensor{\Sigma} &= -p\tensor{I} + \tensor{\tau}_\mathrm{am} + \tensor{\tau}_\mathrm{sc}, \qquad \qquad
\tensor{\tau}_\mathrm{am} = G \left(\frac{1}{\zeta}\frac{1}{1-x} \tensor{c} - \tensor{I}\right),
\qquad \qquad \tensor{\tau}_\mathrm{sc} = 3G \left(\tensor{S} + 2\lambda_\mathrm{sc}\tensor{U}\right), \\
\lambda_\mathrm{am} \; \overset{\triangledown}{\tensor{c}} &= - \Bigg( (1-\alpha)\tensor{I} + \frac{\alpha}{\zeta} \frac{1}{1-x} \tensor{c}\Bigg) \Bigg( \tensor{c} - \zeta (1-x) \tensor{I} \Bigg), \qquad \qquad \lambda_\tn{am} = \lambda (1-x)^2, \qquad \qquad \lambda = \frac{\mu}{G}, \\
\lambda_\mathrm{sc} \; \overset{\triangledown}{\tensor{S}} &= -\sigma \tensor{S} + \frac{1}{3}\lambda_\mathrm{sc} \left(\nabla \vece{v} + \left(\nabla \vece{v}\right)^T \right) - 2\lambda_\mathrm{sc}\tensor{U}, \hspace*{2.4cm} \lambda_\tn{sc} = \delta \lambda \exp(Fx), \\
K(T,a) &= K_\mathrm{max} \exp(-4 \, \tn{ln}(2) \frac{(T-T_\tn{max})^2}{(\Delta T)^2} + 2 \frac{\xi}{G} a),
\end{align*}
\end{subequations}
supplemented with appropriate inlet and outlet boundary conditions.
\end{system}
The fiber geometry is specified by the kinematic boundary condition on the free surface $\Gamma_\mathrm{fr}$ with unit outer normal vector $\vece{n}$.
In System~\ref{system_3d}, $\tensor{\Sigma}$ denotes the stress tensor with contributions from pressure $p$, amorphous extra stress $\tensor{\tau}_\tn{am}$ and semi-crystalline extra stress $\tensor{\tau}_\tn{sc}$ with unit tensor $\tensor{I}$ and operator $\overset{\triangledown}{\cdot}$ for the upper-convected derivative. The operator $\langle \cdot , \cdot \rangle_F$ denotes the Frobenius inner product.
The body force $\vece{g}$ comes from gravity, the surface forces $\vece{f}_\mathrm{st}$ and $\vece{f}_\mathrm{air}$ are due to surface tension and the surrounding airflow. The heat transfer on the lateral surface depends on the difference between fiber temperature and air temperature $T_\mathrm{air}$ and the heat transfer coefficient $\alpha_T$. The mass density $\rho(T,x)$, dynamic viscosity $\mu(T)$, specific heat capacity $c_\tn{p}(T,x)$, specific latent heat of crystallization $\Delta H_\mathrm{f}(T,x)$ and thermal conductivity $C(T,x)$ might be modeled as temperature and crystallinity-dependent functions. The amorphous and semi-crystalline relaxation times are denoted by $\lambda_\mathrm{am}$ and $\lambda_\mathrm{sc}$, respectively.
The energy-driven crystallization rate $K(T,a)$ contains the maximum crystallization rate $K_\tn{max}$, temperature of maximum crystallization rate $T_\tn{max}$, temperature half-width $\Delta T$ and flow-enhanced crystallization (FEC) parameter $\xi$.
In the amorphous constitutive equation, $\alpha$ is the Giesekus mobility parameter and $\zeta=N_0 l^2 / 3$ with $N_0$ number of flexible statistical links of length $l$ of one polymer chain. In the semi-crystalline constitutive equation, $\sigma$ is the anisotropic drag coefficient and $\tensor{U}$ the closure approximation tensor (cf.~Appendix \ref{appendix:closure}).
Furthermore, $\Phi_\infty$ denotes the ultimate degree of crystallization, $G$ the melt shear modulus, $\delta$ and $F$ are model parameters for the semi-crystalline relaxation time. For details on the modeling we refer to \cite{doufas_simulation_2000-1,doufas_simulation_2000,shrikhande_modified_2006}.

\section{Dimensionally reduced fiber models} \label{sec:reducedModels}

A fiber is a slender body, whose radius is typically orders of magnitude smaller than its length. Hence, fiber spinning in a surrounding air flow is a multiphase-multiscale problem in a complex geometry, whose direct numerical simulation is in general computationally not possible due to the high complexity.
A common modeling approach consists thus of the separate consideration of fiber dynamics and air flow, a dimensional reduction of the fiber model with, e.g., slenderbody-theory and then a coupling of fiber and air models via appropriate exchange functions. Embedded into an iterative coupling algorithm, this makes the simulation of fiber spinning with fiber-air-interactions for industrial setups with up to thousands of fibers possible, see, e.g., \cite{ettmueller_ecmi_2024}.
The success of this procedure, however, depends on the quality of the underlying simplified fiber model. The dimensionally reduced model should capture all physically relevant effects, while being cheap and fast to evaluate with a reasonable computational effort at the same time.

Consider an uniaxial radially symmetric spinning setup in cylindrical coordinates with origin in the fiber center at the point of maximum die swell, as sketched in Fig.~\ref{fig:sketchMeltSpinning}. The symmetry allows the restriction to a two-dimensional cutting plane spanned by the orthonormal basis $\lbrace \vece{a}_z, \vece{a}_r \rbrace$, where $\vece{a}_z$ points from the inlet to the outlet in direction of the symmetry axis.
Then, the fiber coordinates are in $\mathcal{Q}_{cut} = \lbrace (s,r) \in \mathbb{R}^2 \,|\,  r \in [0,R(s)], s \in [0,L] \rbrace$ with radius function $R$ and fiber length $L$. In addition, the model equations of System~\ref{system_3d} can be dimensionally reduced by cross-sectional averaging. In this section we present and discuss two models that combine cross-sectionally averaged equations with radially resolved equations, the stress-resolved 1D2D model and the stress-averaged 1D2D model. The well-known cross-sectionally averaged 1D fiber model \cite{doufas_simulation_2000-1, ettmuller_asymptotic_2021, shrikhande_modified_2006} completes here our model hierarchy.
In accordance to the mentioned literature, we assume at the inlet a constant temperature profile $T_\mathrm{in}$ as well as zero crystallization $x_\text{in}\equiv 0$ and zero stored energy $a_\text{in}\equiv 0$.
Note that all fiber models in this section are considered in non-dimensional form. For more information on their derivation and non-dimensionalization we refer to Appendix~\ref{appendix:derivation}. The reference values used for the non-dimensionalization and the resulting dimensionless characteristic numbers are listed in Table~\ref{tab:nondim}.

\begin{table}[t]
\centering
\begin{tabular}{|l r@{\,}c@{\,}l l|}
\hline
\multicolumn{5}{|l|}{\textbf{Reference values}}							\\
Description					& \multicolumn{3}{l}{Formula}					& Unit	\\
\hline
Length						& $s_0$			&$=$& $L$						& m		\\
Cross-sectional area		& $A_0$			&$=$& $A_\text{in}=\pi D_\text{in}^2/4$			& $\tn{m}^2$ \\
Diameter					& $D_0$			&$=$& $\sqrt{A_0}$				& m		\\
Radius						& $R_0$			&$=$& $D_0$						& m 	\\
Velocity					& $u_0$			&$=$& $u_\text{in}$					& m/s	\\
Temperature					& $T_0$			&$=$& $\mathfrak{I}(T_\text{in})$					& K		\\
Conformation				& $c_0$			&$=$& $\zeta$					& $\tn{m}^2$ \\
Free energy					& $a_0$			&$=$& $G$						& Pa	\\
Density						& $\rho_0$		&$=$& $\rho(T_0)$			& kg/$\tn{m}^3$ \\
Viscosity					& $\mu_0$		&$=$& $\mu(T_0)$				& Pa s 	\\
Specific heat capacity				& $c_{\mathrm{p},0}$		&$=$& $c_\mathrm{p}(T_0,0)$			& J/(kg K) \\
Heat of fusion				& $\Delta H_{\mathrm{f},0}$ &$=$& $\Delta H_\mathrm{f}(T_0,0)$	& J/kg	\\
Thermal conductivity 		& $C_0$ &$=$& $C(T_0,0)$ & W/(m K) \\
Heat transfer coefficient	& $\alpha_{\mathrm{T},0}$			&$=$& $\alpha_{\mathrm{T},\mathrm{in}}$		& W/($\tn{m}^2$ K) \\
Stress						& $\tau_0$		&$=$& $\mu_0 u_0 / s_0$			& Pa	\\
Outer force					& $f_0$			&$=$& $\rho_0 A_0 u_0^2 / s_0$	& N/m \\
Relaxation time				& $\lambda_0$	&$=$& $\mu_0/G$				& s		\\
Crystallization rate		& $K_0$			&$=$& $K_\text{max}$					&1/s	\\
\hline
\multicolumn{1}{l}{} \\
\hline
\multicolumn{5}{|l|}{\textbf{Dimensionless numbers}}							\\
Description				& \multicolumn{3}{l}{Formula}							&\\
\hline
Slenderness				& $\epsilon$		&$=$& $ D_0/s_0$					&\\
Reynolds				& $\tn{Re}$			&$=$& $\rho_0 u_0 s_0/\mu_0$		&\\
Froude					&$\tn{Fr}$			&$=$& $u_0/\sqrt{s_0\,g}$			&\\
Capillary				&$\tn{Ca}$			&$=$& $\mu_0 u_0 / \gamma$		&\\
Peclet					&$\tn{Pe}$			&$=$& $c_{\tn{p},0} \rho_0 u_0 R_0 / C_0$		& \\
Stanton					&$\tn{St}$			&$=$& $\alpha_{\mathrm{T},0}/(c_{\mathrm{p},0} \rho_0 u_0)$	&\\
Eckert					&$\tn{Ec}$			&$=$& $u_0^2/(c_{\mathrm{p},0}T_0)$	&\\
Jakob					&$\tn{Ja}$			&$=$& $c_{\mathrm{p},0}T_0/ (\Delta H_{\mathrm{f},0})$&\\
Deborah					&$\tn{De}$			&$=$& $\lambda_0 u_0/s_0$				&\\
Damk\"ohler				&$\tn{Da}$	&$=$& $K_0 s_0 / u_0$				&\\
Draw ratio					&$\tn{Dr}$			&$=$& $u_{\mathrm{out}}/u_\mathrm{in}$ 				&\\
\hline
\end{tabular}
\vspace{.3cm}
\caption{Reference values used for non-dimensionalization and resulting dimensionless characteristic numbers. The indices $_\text{in}$ and $_\text{out}$ indicate boundary data at inlet and outlet, respectively.}
\label{tab:nondim}
\end{table}

\subsection{Stress-resolved one-two-dimensional model} \label{subsec:stress-resolved}
The stress-resolved 1D2D model presented in System~\ref{system_fullradial} corresponds to the fiber model proposed by Doufas et al. in \cite{doufas_two-dimensional_2001}. Mass and momentum balances are cross-sectionally averaged, whereas energy balance and all equations related to the extra stress tensor components are radially resolved. Thus, the mass flow is constant. We structure the fiber model into three parts according to the type of differential equations: 1) the two ordinary differential equations for the scalar-valued axial velocity $u$ and its derivative $\omega=\partial_s u$ in \eqref{eqFull:onetwo_averaged} result from the cross-sectionally averaged momentum balance; 2) the energy balance becomes a partial differential advection-diffusion equation for the temperature $T$ in \eqref{eqFull:onetwo_diffusion}; and 3) the differential equations for the conformation tensor components $c_{zz}$ and $c_{rr}$, the scaled orientational tensor component $S$, crystallinity $x$ and stored free energy $a$ in \eqref{eqFull:onetwo_material} have a parametrically radial dependence due to temperature, but do not contain any radial derivatives. Note that this dependency is a result of the performed transformation from the unknown fiber domain $\Omega_{cut}$ with free boundary to the unit square $(s,r)\in [0,1]^2$ which eliminated all terms with radial component and respective derivatives.
The differences in System~\ref{system_fullradial} compared to the original model in \cite{doufas_two-dimensional_2001} come from a different non-dimensionalization, a scaling of $S$ and the choice of an energy-driven crystallization rate (cf.\ Appendix~\ref{appendix:derivation}). For the boundary conditions we follow \cite{ettmuller_asymptotic_2021}, as discussed below. The non-dimensional stress-resolved 1D2D fiber model for the unknowns $\bs{y} = (u, \omega, T, c_{zz}, c_{rr}, S, x, a)$ is given by System~\ref{system_fullradial}.
\begin{system}[Stress-resolved one-two-dimensional model] \label{system_fullradial} ~\newline
Cross-sectionally averaged equations $s \in [0,1]$:
\begin{subequations}
\begin{align} \label{eqFull:onetwo_averaged}
\begin{split}
\partial_s u &= \omega, \\
\delta \, \avgUnit{\abkU(\bs{y};\delta)} \partial_s \omega &= \Omega_0(\bs{y}) + \delta \; \Omega_1(\bs{y},\delta),
\end{split}
\end{align}
with boundary conditions
\begin{align*}
u\big|_{s=0} = 1, \qquad\qquad u\big|_{s=1} = \tn{Dr}, \qquad\qquad \Omega_0(\bs{y}\big|_{s=0}) = 0.
\end{align*}
Advection-diffusion equation $(s,r) \in [0,1]^2$:
\begin{align} \label{eqFull:onetwo_diffusion}
 c_\mathrm{p}(\avgUnit{T},\avgUnit{x}) \, \rho u\, \partial_s T - \frac{1}{\epsilon \tn{Pe}} \frac{C(\avgUnit{T},\avgUnit{x})}{R^2 r} \partial_r \left(r \partial_r T \right) = \frac{1}{\tn{Ja}} \Phi_\infty\Delta H_\mathrm{f}(\avgUnit{T},\avgUnit{x})\, \rho u  \,\partial_s x,
\end{align}
with boundary conditions
\begin{align*}
T\Big|_{s=0} = 1, \qquad\qquad \partial_r T\Big|_{r=0} = 0, \qquad\qquad \partial_r T\Big|_{r=1} = - \tn{St}\;\tn{Pe} \left(\frac{\alpha_\mathrm{T} R}{C(\avgUnit{T},\avgUnit{x})} \left(T - T_\mathrm{air}\right)\right) \Bigg|_{r=1}.
\end{align*}
Parametrically radially dependent equations $(s,r) \in [0,1]^2$:
\begin{align} \label{eqFull:onetwo_material}
\begin{split}
\partial_s c_{zz} &= 2\frac{c_{zz}}{u} \omega - \frac{1}{\tn{De}} \frac{1}{\lambda(T) u} \bigg((1-\alpha) + \alpha \frac{c_{zz}}{1-x} \bigg) \bigg(\frac{c_{zz}}{(1-x)^2} - \frac{1}{1-x} \bigg), \\
\partial_s c_{rr} &= - c_{rr} \left(\frac{\omega}{u} + \frac{\partial_s \rho}{\rho} \right) - \frac{1}{\tn{De}} \frac{1}{\lambda(T) u} \bigg((1-\alpha) + \alpha \frac{c_{rr}}{1-x} \bigg) \bigg(\frac{c_{rr}}{(1-x)^2} - \frac{1}{1-x} \bigg), \\
\delta \,\partial_s S &= \mathcal{S}_0(\bs{y}) + \delta \; \mathcal{S}_1(\bs{y},\delta), \\
\partial_s x &= \tn{Da} \, K(T,a) \, \frac{1-x}{u}, \\
\partial_s a &= - \frac{1}{\tn{De}} \frac{a}{\lambda(T) u (1-x)^2} + \frac{1}{1-x} \left( c_{zz} \frac{\omega}{u} - c_{rr} \left( \frac{\omega}{u} + \frac{\partial_s \rho}{\rho} \right) \right),
\end{split}
\end{align}
with boundary conditions
\begin{align*}
c_{zz}\big|_{s=0}+2c_{rr}\big|_{s=0} &= 3, &c_{zz}\big|_{s=0} &= c_\mathrm{in}=const, &\mathcal{S}_0(\bs{y}\big|_{s=0}) &= 0,\\
x\big|_{s=0} &= 0, &a\big|_{s=0} &= 0.
\end{align*}
Abbreviations:
\begin{align*}
\avgUnit{f} &= 2 \int_0^1 f(s,r)r\dr, \qquad \qquad \abkU(\bs{y};\delta) = 6 \lambda(T) \exp(Fx) \left(\closureFzU(\delta S) - \closureFrU(\delta S)\right), \\
& \hspace{4.47cm}\abkRho(\bs{y};\delta) = 6 \lambda(T) \exp(Fx) \left(\closureFzRho(\delta S) - \closureFrRho(\delta S)\right), \\
\Omega_0(\bs{y}) &= \tn{Re}\, \rho u  \omega - \tn{Re} \, \rho u  f_\tn{air} - \frac{\tn{Re}}{\tn{Fr}^2} \rho + \frac{\sqrt{\pi}}{2} \frac{1}{\epsilon \tn{Ca}} \sqrt{\rho u} \left( \frac{\omega}{u} + \frac{\partial_s \rho}{\rho} \right) \\
& \quad + \frac{1}{\tn{De}} \left( \avgUnit{\frac{c_{zz}-c_{rr}}{1-x}} \left( \frac{\omega}{u} + \frac{\partial_s \rho}{\rho} \right) - \avgUnit{\partial_s \left(\frac{c_{zz}-c_{rr}}{1-x}\right)}\right) - \frac{9}{2}\frac{1}{\tn{De}} \avgUnit{\mathcal{S}_0(\bs{y})}, \\
\Omega_1(\bs{y},\delta) &= \left(\omega \avgUnit{\abkU(\bs{y};\delta)} + \frac{9}{2}\frac{1}{\tn{De}} \avgUnit{S} \right) \left( \frac{\omega}{u} + \frac{\partial_s \rho}{\rho} \right) - \omega \avgUnit{\partial_s \abkU(\bs{y};\delta)} - \frac{9}{2}\frac{1}{\tn{De}} \avgUnit{\mathcal{S}_1(\bs{y},\delta)} \\
& \quad + \avgUnit{\abkRho(\bs{y};\delta)} \partial_s \left( u \frac{\partial_s \rho}{\rho} \right) + \avgUnit{\partial_s \abkRho(\bs{y};\delta)} u \frac{\partial_s \rho}{\rho} + \avgUnit{\abkRho(\bs{y};\delta)} u \frac{\partial_s \rho}{\rho} \left( \frac{\omega}{u} + \frac{\partial_s \rho}{\rho} \right), \\
\mathcal{S}_0(\bs{y}) &= - \frac{\sigma}{\tn{De}} \frac{S}{\lambda(T) \exp(Fx)u } + \frac{2}{5} \frac{\omega}{u}, \qquad \qquad \mathcal{S}_1(\bs{y};\delta) = 2 (S - \closureFzUTilde(S;\delta)) \frac{\omega}{u} - 2 \closureFzRhoTilde(S;\delta) \frac{\partial_s \rho}{\rho},\\
\closureFzUTilde(S;\delta) &= -\frac{81}{8}\delta^4 S^5 + \frac{675}{56}\delta^3 S^4 - \frac{36}{35}\delta^2 S^3 - \frac{9}{10}\delta S^2 + \frac{11}{14}S, \\
\closureFzRhoTilde(S;\delta) &= -\frac{27}{8}\delta^4 S^5 + \frac{153}{56}\delta^3 S^4 + \frac{9}{14}\delta^2 S^3 + \frac{1}{14}S.
\end{align*}
\end{subequations}
\end{system}

Cross-sectional averaging is described by the integral operator $\mathfrak{I}$. The radius function is given by $R(s)=1/\sqrt{\pi \rho u(s)}$. The density $\rho$ may depend on the cross-sectionally averaged temperature and crystallinity, i.e., $\rho = \rho(\avgUnit{T},\avgUnit{x})$ (cf.\ Remark~\ref{remark:densityAverage} in Appendix \ref{appendix:derivation}). The scalar-valued axial drag force is denoted by $f_\mathrm{air}$, for the closure approximations $\mathcal{U}_{z,u}$, $\mathcal{U}_{r,u}$, $\mathcal{U}_{z,\rho}$, $\mathcal{U}_{r,\rho}$ we refer to Appendix~\ref{appendix:closure}. Note that in $\Omega_0$, $\Omega_1$ and $\mathcal{S}_1$ the notation of the derivative is used as abbreviation for the respective expression in terms of the variables $\bs{y}$. 

The boundary conditions can be deduced from the three-dimensional description (System~\ref{system_3d}) -- with one exception. Due to the radial symmetry, we obtain straightforward the conditions for the temperature $T$ at the inlet, the symmetry axis and the fiber surface, cf.\ \eqref{eqFull:onetwo_diffusion}. Note that a constant temperature profile at the inlet is underlying. The zero-inlet profiles for $x$ and $a$ are also handed over. The inlet relation for the conformation tensor components, $c_{zz}\big|_{s=0}+2c_{rr}\big|_{s=0} = 3$, results from the respective trace condition on the tensor and emphasizes the viscous fiber behavior near the nozzle. The inlet and outlet conditions for the axial (cross-sectionally averaged) velocity $u$ correspond to their three-dimensional analogon for inflow and take-up. For $\omega$ and $S$ we apply the asymptotically-justified inlet conditions from \cite{ettmuller_asymptotic_2021}, i.e.,
$\Omega_0(\bs{y}\big|_{s=0}) = 0$ and $\mathcal{S}_0(\bs{y}\big|_{s=0}) =0$,
to avoid the artificial boundary layers that occurred in the original model \cite{doufas_two-dimensional_2001}, where instead the velocity derivative obtained from the amorphous fiber model and a vanishing orientational tensor component were used at the point of onset of crystallization. In sum, this set of boundary conditions is not enough to close the equations \eqref{eqFull:onetwo_averaged} and \eqref{eqFull:onetwo_material}, but one further (scalar-valued) condition is required.

The need of a further closure condition comes from the unbalanced treatment of the momentum balance and the microstructural equations.
The cross-sectionally averaging of the momentum balance yields two 1D equations for $u$, $\omega$ and three 1D boundary conditions (i.e., inlet and outlet velocity as well as algebraic relation $\Omega_0$).
The microstructural equations that are not averaged but radially resolved result in three 2D ---parametrically radially dependent--- equations for $c_{zz}$, $c_{rr}$ and $S$, but only two 2D boundary conditions (i.e., relation for conformation tensor components as well as for $\mathcal{S}_0$). To close the fiber model, Doufas et al.\ \cite{doufas_two-dimensional_2001} proposed a constant inlet profile of the conformation tensor component $c_{zz}$, i.e., 
\begin{align} \label{eq:bcConstDoufas}
c_{zz}\big|_{s=0} = c_\mathrm{in}= const.
\end{align}
They argued that \eqref{eq:bcConstDoufas} is implied by the homogeneity in temperature at the inlet.
Although this choice might be a reasonable assumption, we point out that it is not an implication from the 3D model equations. There is a direct relationship between the inlet temperature profile and the derivatives of the conformation tensor components, but no relationship between the inlet temperature profile and conformation tensor components themselves. Thus, in this work, we use \eqref{eq:bcConstDoufas} as closure condition in System~\ref{system_fullradial} and investigate its approximation quality in comparison with 3D reference simulations (cf.\ Section~\ref{sec:compareFenics}).

\subsection{Stress-averaged one-two-dimensional model}
The stress-resolved 1D2D model requires an additional radial profile assumption as discussed in Section~\ref{subsec:stress-resolved}
As an alternative, we propose a stress-averaged 1D2D model where the microstructural equations are cross-sectionally averaged and only temperature and crystallinity are radially resolved. A similar averaging strategy was applied in \cite{ettmuller_productIntegration_2023} to derive a viscous 1D2D model.
The non-dimensional stress-averaged 1D2D model is given by System~\ref{system_intermediate}. We structure the model accordingly to System~\ref{system_fullradial} into three parts with respect to the type of differential equations. Since the equation for the crystallinity is only parametrically radially dependent, no additional radial profile assumption is required. If not stated otherwise, all abbreviations and functions are taken from Section \ref{subsec:stress-resolved}.

\begin{system}[Stress-averaged one-two-dimensional model] \label{system_intermediate} ~\newline
Cross-sectionally averaged equations $s \in [0,1]$:
\begin{subequations}
\begin{align} \label{eqInter:onetwo_averaged}
\begin{split}
\partial_s u &= \omega, \\
\delta \, \avgUnit{\abkU(\bs{y};\delta)} \partial_s \omega &= \Omega_0(\bs{y}) + \delta \; \Omega_1(\bs{y},\delta), \\
\partial_s c_{zz} &= 2\frac{c_{zz}}{u} \omega - \frac{1}{\tn{De}} \avgUnit{\frac{1}{\lambda(T) u} \bigg((1-\alpha) + \alpha \frac{c_{zz}}{1-x} \bigg) \bigg(\frac{c_{zz}}{(1-x)^2} - \frac{1}{1-x} \bigg)}, \\
\partial_s c_{rr} &=  - c_{rr} \left(\frac{\omega}{u} + \frac{\partial_s \rho}{\rho} \right) - \frac{1}{\tn{De}} \avgUnit{\frac{1}{\lambda(T) u} \bigg((1-\alpha) + \alpha \frac{c_{rr}}{1-x} \bigg) \bigg(\frac{c_{rr}}{(1-x)^2} - \frac{1}{1-x} \bigg)}, \\
\delta \,\partial_s S &= \mathcal{S}_0(\bs{y}) + \delta \; \mathcal{S}_1(\bs{y},\delta), \\
\partial_s a &= - \frac{1}{\tn{De}} \avgUnit{\frac{1}{\lambda(T)(1-x)^2}} \frac{a}{u} + \avgUnit{\frac{1}{1-x}} \left( c_{zz} \frac{\omega}{u} - c_{rr} \left( \frac{\omega}{u} + \frac{\partial_s \rho}{\rho} \right) \right),
\end{split}
\end{align}
with boundary conditions
\begin{align*}
u\big|_{s=0}&= 1, \quad\qquad u\big|_{s=1} = \tn{Dr}, \,\quad\qquad  \Omega_0(\bs{y}\big|_{s=0}) = 0,\\[.2cm]
 c_{zz}\big|_{s=0}+2c_{rr}\big|_{s=0}&= 3, \quad \qquad a\big|_{s=0} = 0, \qquad\qquad \mathcal{S}_0(\bs{y}\big|_{s=0}) = 0.
\end{align*}
Advection-diffusion equation $(s,r) \in [0,1]^2$:
\begin{align} \label{eqInter:onetwo_diffusion}
c_\mathrm{p}(\avgUnit{T},\avgUnit{x}) \,\rho u\, \partial_s T - \frac{1}{\epsilon \tn{Pe}} \frac{C(\avgUnit{T},\avgUnit{x})}{R^2 r} \partial_r \left(r \partial_r T \right) = \frac{1}{\tn{Ja}} \Phi_\infty\Delta H_\mathrm{f}(\avgUnit{T},\avgUnit{x})\, \rho u\, \partial_s x,
\end{align}
with boundary conditions
\begin{align*}
T\Big|_{s=0} = 1, \quad\qquad \partial_r T\Big|_{r=0} = 0, \quad\qquad \partial_r T\Big|_{r=1} = - \tn{St}\;\tn{Pe} \left(\frac{\alpha_\mathrm{T}R}{C(\avgUnit{T},\avgUnit{x})} \left(T - T_\mathrm{air}\right)\right) \Bigg|_{r=1}.
\end{align*}
Parametrically radially dependent equation $(s,r) \in [0,1]^2$:
\begin{align} \label{eqInter:onetwo_material}
\partial_s x &= \tn{Da} \, K(T,a) \, \frac{1-x}{u},
\end{align}
\end{subequations}
with boundary condition
\begin{align*}
&x\big|_{s=0} = 0.
\end{align*}
Abbreviations:
\begin{align*}
\mathcal{S}_0(\bs{y}) &= - \frac{\sigma}{\tn{De}} \avgUnit{\frac{1}{\lambda(T) \exp(Fx)}} \frac{S}{u} + \frac{2}{5} \frac{\omega}{u}.
\end{align*}
\end{system}

\begin{remark}
There are different ways for averaging the equations for the components of the conformation and orientational tensors in \eqref{eqInter:onetwo_averaged}. In the presented approach we take the cross-sectional average of the right-hand side function and apply basic properties of the integral operator to arrive at the equations shown above.
Alternatively, temperature and crystallinity, or specific individual terms on the right-hand side such as the relaxation time could be averaged first. However, in such a procedure, the non-linearities are not correctly averaged.
\end{remark}

\subsection{Averaged one-dimensional model}
System~\ref{system_averaged} shows the well known cross-sectionally averaged 1D fiber model, cf.\ \cite{doufas_simulation_2000-1, ettmuller_asymptotic_2021, shrikhande_modified_2006}.
Proceeding in the model hierarchy, it results from averaging the advection-diffusion equation for the temperature \eqref{eqInter:onetwo_diffusion} over the fiber cross-section and using the radius function $R(s)=1/\sqrt{\pi \rho u(s)}$, \begin{align*}
\partial_s \avgUnit{T} =  -2\sqrt{\pi} \frac{\tn{St}}{\epsilon} \frac{\alpha_\mathrm{T}}{c_\mathrm{p}(\avgUnit{T},\avgUnit{x}) \sqrt{\rho u}} \left( T\Big|_{r=1} - T_\mathrm{air} \right) + \frac{1}{\tn{Ja}} \frac{\Phi_\infty \Delta H_\mathrm{f}(\avgUnit{T},\avgUnit{x})}{c_\mathrm{p}(\avgUnit{T},\avgUnit{x})} \partial_s \avgUnit{x}.
\end{align*}
Then the equation for the crystallinity $x$ only depends on the averaged temperature and thus becomes also one-dimensional.

\begin{system}[Averaged one-dimensional model] \label{system_averaged} ~\newline
Cross-sectionally averaged equations $s \in [0,1]$:
\begin{align*}
\deriv{u}									&= \omega, \\
\delta \, \abkU(\bs{y};\delta) \deriv{\omega} 	&= \Omega_0(\bs{y}) + \delta \; \Omega_1(\bs{y};\delta), \\
\deriv{T}				&= - 2 \sqrt{\pi} \frac{\tn{St}}{\epsilon} \frac{\alpha_\mathrm{T}}{c_\mathrm{p}(T,x) \sqrt{\rho u }} (T-T_\mathrm{air}) + \frac{1}{\tn{Ja}} \frac{\Phi_{\infty} \Delta H_\mathrm{f}(T,x)}{c_\mathrm{p}(T,x)} \partial_s x, \\
\deriv{c_{zz}}			&= 2\frac{c_{zz}}{u} \omega- \frac{1}{\tn{De}} \frac{1}{ \lambda(T) u}  \bigg(1-\alpha + \alpha \frac{c_{zz}}{1-x}  \bigg) \bigg(\frac{c_{zz}}{(1-x)^2} - \frac{1}{1-x} \bigg), \\
\deriv{c_{rr}}			&=  - c_{rr} \left(\frac{\omega}{u} + \frac{\partial_s \rho}{\rho} \right) - \frac{1}{\tn{De}} \frac{1}{\lambda(T) u } \bigg(1-\alpha + \alpha \frac{c_{rr}}{1-x} \bigg) \bigg(\frac{c_{rr}}{(1-x)^2} - \frac{1}{1-x} \bigg), \\
\delta \,\deriv{S} 							&= \mathcal{S}_0(\bs{y}) + \delta \; \mathcal{S}_1(\bs{y};\delta), \\
\deriv{x} 				&= \tn{Da}\,K(T,a)\,\frac{1-x}{u}, \\
\deriv{a}				&= - \frac{1}{\tn{De}}\frac{a}{\lambda(T) u  (1-x)^2} + \frac{1}{1-x} \left( c_{zz} \frac{\omega}{u} - c_{rr} \left( \frac{\omega}{u} + \frac{\partial_s \rho}{\rho} \right) \right),
\end{align*}
with boundary conditions
\begin{align*}
u\big|_{s=0}&= 1, \quad\qquad u\big|_{s=1} = \tn{Dr}, \,\quad\qquad  \Omega_0(\bs{y}\big|_{s=0}) = 0,\\[.2cm]
 c_{zz}\big|_{s=0}+2c_{rr}\big|_{s=0}&= 3, \quad \qquad a\big|_{s=0} = 0, \qquad\qquad \mathcal{S}_0(\bs{y}\big|_{s=0}) = 0\\
 T\big|_{s=0}&= 1, \quad \qquad   x\big|_{s=0} = 0.
 \end{align*}
\end{system}

\section{Approximation quality and computational effort} \label{sec:compareFenics}

\begin{figure}[t]
\centering
\includegraphics[width=0.29\textwidth]{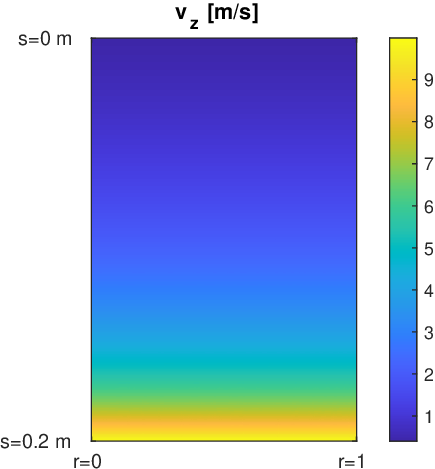}
\includegraphics[width=0.3\textwidth]{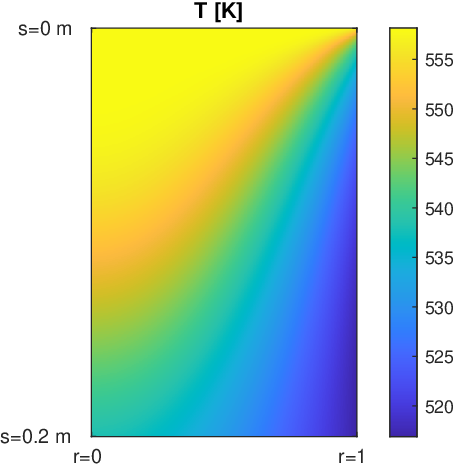}
\includegraphics[width=0.3\textwidth]{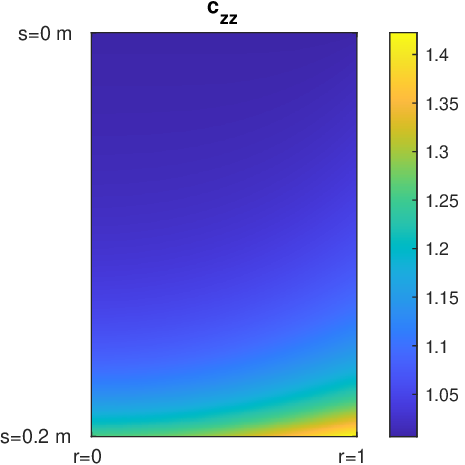} \\
\includegraphics[width=0.3\textwidth]{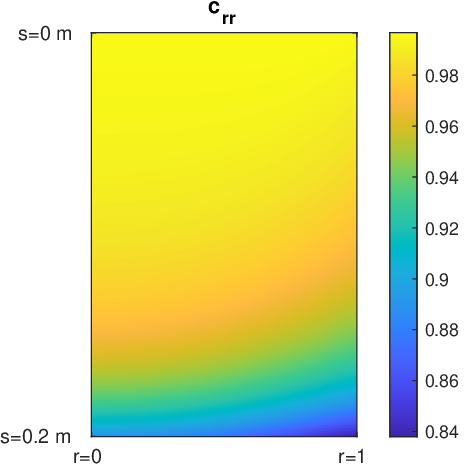}
\includegraphics[width=0.3\textwidth]{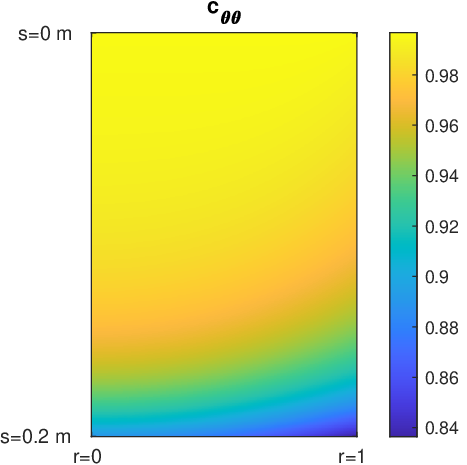}
\includegraphics[width=0.3\textwidth]{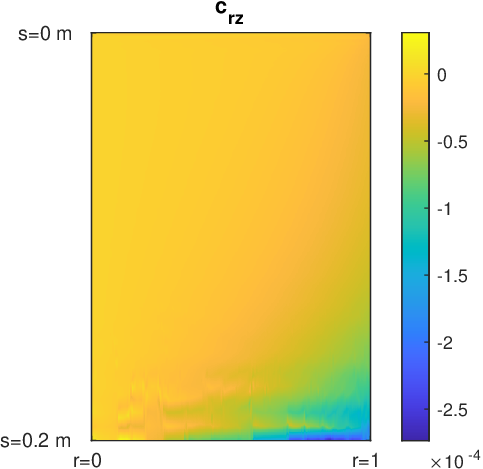}
\caption{Fiber quantities of the 3D reference model in cylindrical coordinates.}
\label{fig:solutionFenics}
\end{figure}

\begin{table}[b]
\centering
\begin{tabular}{|l l l l|}
\hline
\multicolumn{4}{|l|}{\textbf{Physical, rheological and model parameters}}						\\
Description								& Symbol			& Value		& Unit			\\
\hline
Density									& $\rho$			& 1360 		& kg/$\tn{m}^3$ \\
Specific heat capacity				& $c_\mathrm{p}$	& 1400			& J/(kg K) \\
Thermal conductivity				& $C$					& 0.02			& W/(m K) \\
Shear modulus						& $G$		& $5 \cdot 10^5$	& Pa	\\
Heat transfer coefficient			& $\alpha_\mathrm{T}$ & 77.2685 & W/($\tn{m}^2$ K) \\
Giesekus mobility parameter								& $\alpha$			& 0.5  & -\\
Gravitational acceleration		& $g$		& 9.81		& m/$\tn{s}^2$ \\ 
\hline
\multicolumn{4}{l}{}	\\
\hline
\multicolumn{4}{|l|}{\textbf{Process parameters}}											\\
Description								& Symbol			& Value		& Unit			\\
\hline
Fiber length							& $L$				& 0.2		& m				\\
Nozzle diameter							& $D_\text{in}$			& $3 \cdot 10^{-4}$ & m	\\
Temperature at inlet					& $T_\text{in}$			& 558.15	& K				\\
Velocity at inlet						& $v_\text{in}$			& 0.4	& m/s			\\
Take-up velocity at outlet			& $v_\text{out}$			& 10.0		& m/s			\\
Air temperature							& $T_\mathrm{air}$			& 283.15	& K				\\
\hline
\end{tabular}
\caption{Physical, rheological, process and model parameters for the comparison of the models.}
\label{tab:modelComparison_parameters}
\end{table}

In this section, we investigate the approximation quality and computational effort of the dimensionally reduced models of our model hierarchy compared to the 3D reference. For the numerical computation of the 3D fiber model, we use the open-source finite element software Fenics, which is applied to the weak formulation of System~\ref{system_3d}. The dimensionally reduced models are solved by means of an iteration scheme which computes the solution of the respective ODE and PDE parts iteratively using suitable numerical solvers \cite{ettmueller_ecmi_2024}. Details on the numerical implementation can be found in Appendix~\ref{appendix:numerics}. 

Due to the complexity of the 3D model and the associated high computational effort, we limit the comparison to a purely amorphous scenario in an uniaxial radially symmetric geometry. It is achieved by setting $S = x \equiv 0$ and $\lambda_\mathrm{sc} = 0$ in all models. The equation for the free stored energy $a$ then decouples from the other equations and can thus be also eliminated. 
The body forces are considered to be $\vece{f}_\tn{air}=\vece{f}_\tn{st}=\bs{0}$ in System~\ref{system_3d} and hence $f_\tn{air}=0$ as well as $1/\tn{Ca} \rightarrow 0$ in the remaining systems. The choice of the physical, rheological, process and model parameters are given in Table~\ref{tab:modelComparison_parameters}. 
We take all physical parameters -- except of the dynamic viscosity -- as constants, yielding $\rho=c_\tn{p}=C=\alpha_\mathrm{T}=\Delta H_\mathrm{f} = 1$ in dimensionless form. As the non-linearity of the viscosity contributes significantly to radial differences, we use the temperature-dependent function $\mu(T)$ described in Appendix \ref{appendix:nylon}. The inlet and outlet profiles $T_\mathrm{in}$, $v_\mathrm{in}$ and $v_\mathrm{out}$ are also taken as constant. 
The conformation tensor at the inlet $\tensor{c}_\tn{in}$ in System~\ref{system_3d} is chosen such that it fulfills $c_{zz,\tn{in}} + 2c_{rr,\tn{in}} = 3$, $c_{rr,\tn{in}} = c_{\theta \theta,\tn{in}}$ and $c_{rz,\tn{in}}=c_{r\theta,\tn{in}}=c_{z\theta,\tn{in}}=0$ in cylindrical coordinates.

\begin{figure}[t]
\centering
\hspace*{-4.4cm}
\includegraphics[width=0.32\textwidth]{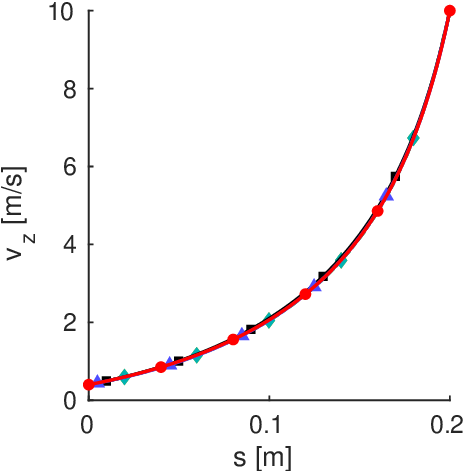}
\includegraphics[width=0.32\textwidth]{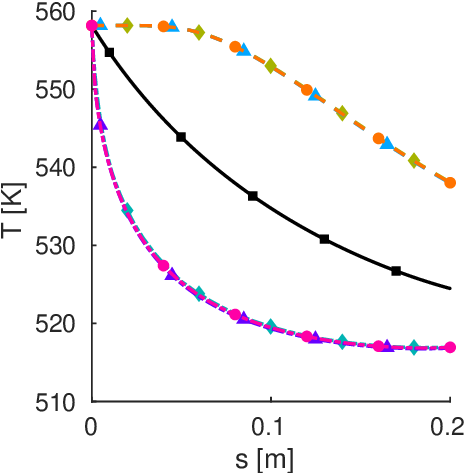} \\[-1.5cm]
\includegraphics[width=0.32\textwidth]{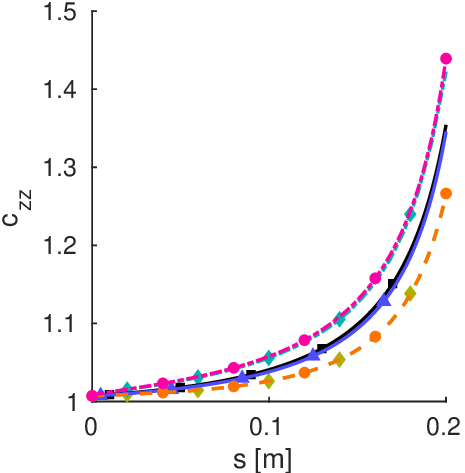}
\includegraphics[width=0.32\textwidth]{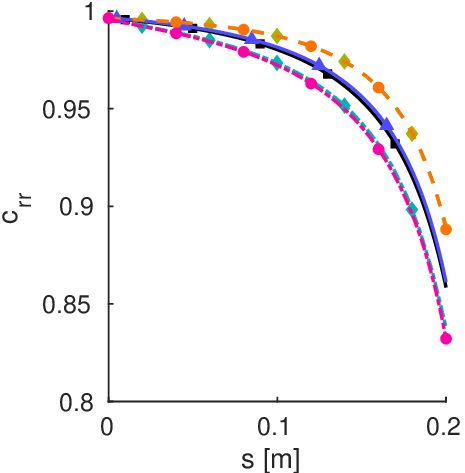} \hspace*{.5cm}
\raisebox{4cm}{\includegraphics[width=0.25\textwidth]{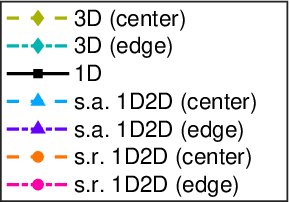}}
\caption{Comparison of models: fiber quantities along the spinline.}
\label{fig:solutionModelComparison}
\end{figure}

The solutions for the fiber quantities of the 3D reference are visualized in Fig.~\ref{fig:solutionFenics}. In agreement with the theoretical results from \cite{henson_thin-filament_1998} the axial velocity is homogeneous over the fiber cross-section, while the temperature forms radial profiles. The fiber has a hot core and becomes colder towards the surface due to the surrounding cooling air. The diagonal entries of the conformation tensor also vary in radial direction with $c_{rr}$ being equivalent to $c_{\theta\theta}$. The off-diagonal component $c_{rz}$ exhibits boundary layers near the outlet position but, in general, can be neglected as its values are five orders of magnitude smaller than the other tensor components. All in all, the conformation tensor is approximately of diagonal shape which is in agreement with the uniaxial flow assumption for the dimensionally reduced models.

\begin{figure}[t]
\centering
\includegraphics[width=0.24\textwidth]{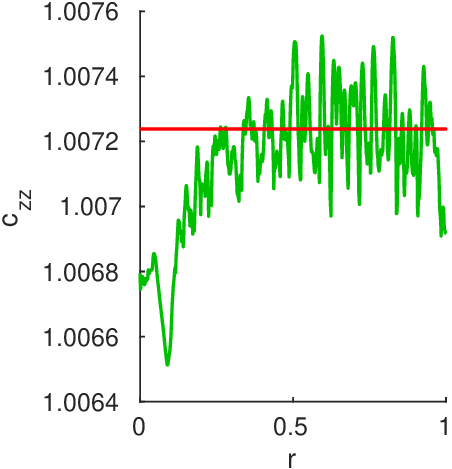}
\includegraphics[width=0.24\textwidth]{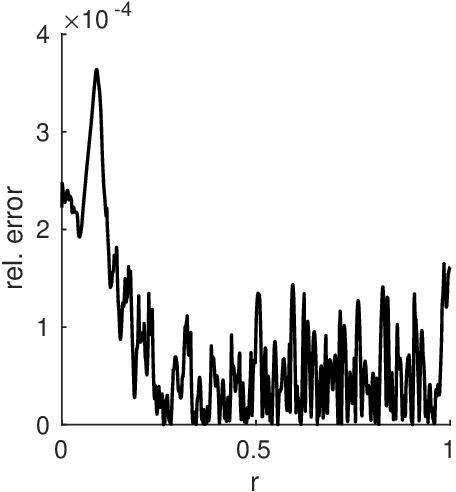}
\hspace{.1cm}
\includegraphics[width=0.24\textwidth]{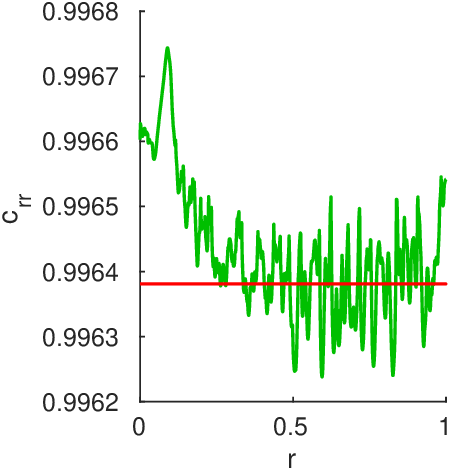}
\includegraphics[width=0.24\textwidth]{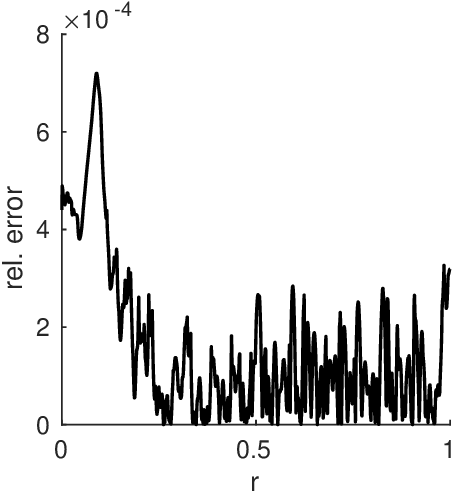}
\caption{Inlet profiles for the conformation tensor components $c_{zz}$ (left) and $c_{rr}$ (right) obtained from the 3D reference (green) and the stress-resolved 1D2D model (green) with the respective relative errors.}
\label{fig:inletModelComparison}
\end{figure}

\begin{table}[b]
\centering
\begin{tabular}{l | c | c | c | c}
					& $u$ & $\avgUnit{T}$ & $\avgUnit{{c}_{zz}}$ & $\avgUnit{{c}_{rr}}$ \\
\hline
\rule{0pt}{1\normalbaselineskip}1D model & $1.3177 \cdot 10^{-2}$ & $2.4573 \cdot 10^{-3}$ & $2.0752 \cdot 10^{-3}$ & $1.0580 \cdot 10^{-3}$ \\
s.a.~1D2D model	& $2.6965 \cdot 10^{-3}$ & $2.2231 \cdot 10^{-4}$ & $2.1542 \cdot 10^{-3}$ & $8.2330 \cdot 10^{-4}$ \\
s.r.~1D2D model			& $2.3795 \cdot 10^{-3}$ & $9.7229 \cdot 10^{-5}$ & $7.3490 \cdot 10^{-4}$ & $3.1983 \cdot 10^{-4}$ 
\end{tabular}
\caption{Relative $L^2$-error between averaged quantities of dimensionally reduced models and 3D reference.}
\label{tab:relL2error}
\end{table}

The outcomes of our model hierarchy (fiber quantities along the spinline) can be compared in Fig.~\ref{fig:solutionModelComparison}.
The stress-resolved 1D2D model demonstrates to be a good approximation of the 3D reference in all aspects. Similarly, the stress-averaged 1D2D model shows strong agreement with the reference in terms of velocity and temperature, but it does not account for radial profiles in the conformation tensor components by construction. The 1D model yields only cross-sectionally averaged information. The investigation of the relative $L^2$-error for the averaged quantities (cf.\ Table~\ref{tab:relL2error}) indicates that the stress-resolved 1D2D model provides the lowest error. Also the stress-averaged 1D2D model performs well, although the error in temperature and conformation tensor components is about one order of magnitude worse. The 1D model shows the highest error. However, with a magnitude of $\mathcal{O}(10^{-3})$ its approximation quality is still very satisfactory. The simplicity of the test case is the primary reason for the good performance of the 1D model. Since the amorphous scenario has no crystallization and semi-crystalline phase contributions, the impact of radial profiles of certain quantities on the overall fiber dynamics is small. However, the accuracy of the approximations improves, the more quantities are resolved radially, but the price to be paid is higher computing times. Whereas the 1D model is solved in about 6 seconds, the stress-averaged and stress-resolved 1D2D models take about 47 and 62 seconds, respectively, with the initialization step for both models requiring about 6 seconds and the radial coupling step about 41 and 56 seconds, respectively.
The computation of the 3D reference model takes 7500 seconds (about 2.1 hours) such that the solving of the dimensionally reduced models is faster by a factor of 120 to 1250 in comparison.

Although the stress-resolved 1D2D model seems to provide the best possible approximation of the reference model, we have to check the suitability of the additional radial profile assumption \eqref{eq:bcConstDoufas}. We investigate the inlet profiles for the tensor components $c_{zz}$ and $c_{rr}$ of both the 3D reference and the stress-resolved 1D2D model, shown in Fig.~\ref{fig:inletModelComparison}. Due to the assumption $c_{zz}|_{s=0}=const$ \eqref{eq:bcConstDoufas} and the prescribed viscous relation between $c_{zz}|_{s=0}+2c_{rr}|_{s=0}=3$, both inlet profiles are constant in the case of the stress-resolved 1D2D model. 
In the reference model, the profiles are non-constant, however only a relative difference of $\mathcal{O}(5\cdot 10^{-4})$ between the maximum and minimum values can be observed.
The relative error at the inlet is thus in the same order of magnitude as the overall approximation quality of the stress-resolved 1D2D model, cf.~Table~\ref{tab:relL2error}. Changing the inlet and outlet velocities does not significantly affect the qualitative differences between the inlet profiles.
This shows the validity of the profile assumption in the case of the amorphous scenario under consideration. Whether \eqref{eq:bcConstDoufas} is a sufficiently reliable assumption for all parameter ranges or even for the two-phase model, remains an open question. In practical applications it may be preferable to consider the stress-averaged 1D2D model, as it does not require additional assumptions about the microstructure at the boundary.

\section{Nylon test case scenario} \label{sec:nylonCase}
This section presents a performance study on the dimensionally reduced models in a practical setting.
We consider a test case for spinning Nylon-66 with a take-up velocity of 1000 $\tn{m}/\tn{min}$, where low to moderate effects of flow-enhanced crystallization (FEC) are expected. 
In addition, to investigate the impact of the flow-enhanced crystallization on the validity on the fiber models, the FEC parameter $\xi$ is artificially varied. All relevant physical, rheological, process and model parameters are listed in Appendix~\ref{appendix:nylon}. 

\begin{figure}[t]
\centering
\includegraphics[width=0.26\textwidth]{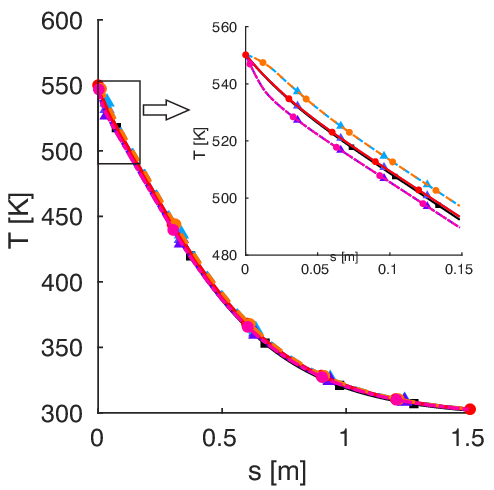}
\includegraphics[width=0.26\textwidth]{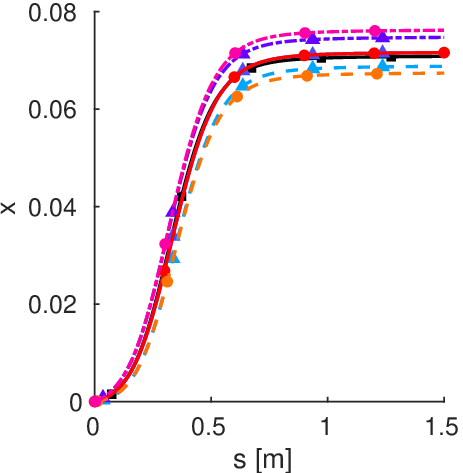}
\includegraphics[width=0.26\textwidth]{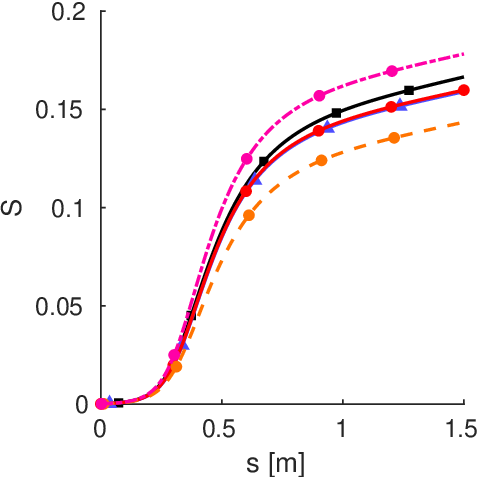}
\hspace*{0.1cm}
\raisebox{1.2cm}{\includegraphics[width=0.18\textwidth]{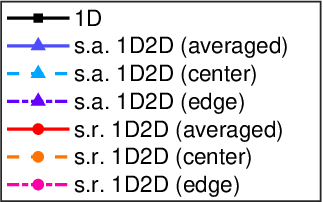}}
\caption{Temperature $T$, crystallinity $x$ and orientational tensor component $S$ along the spinline (cross-sectionally averaged, at fiber center and at surface) for all three models.}
\label{fig:nylonTestCaseAxialSolutions}
\end{figure}

\begin{figure}[t]
\centering
\includegraphics[width=0.26\textwidth]{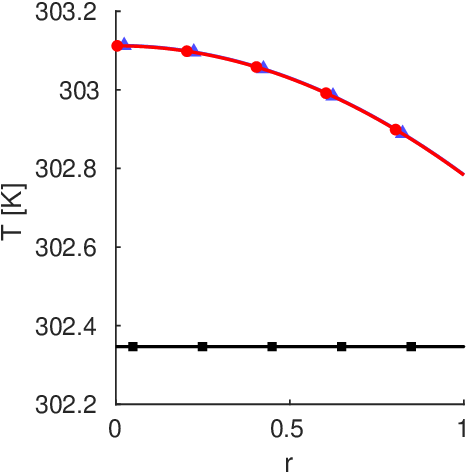}
\includegraphics[width=0.26\textwidth]{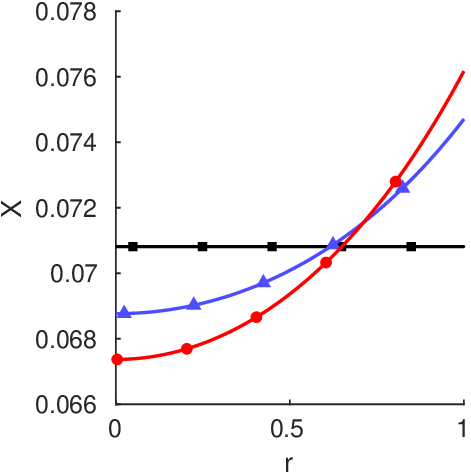}
\includegraphics[width=0.26\textwidth]{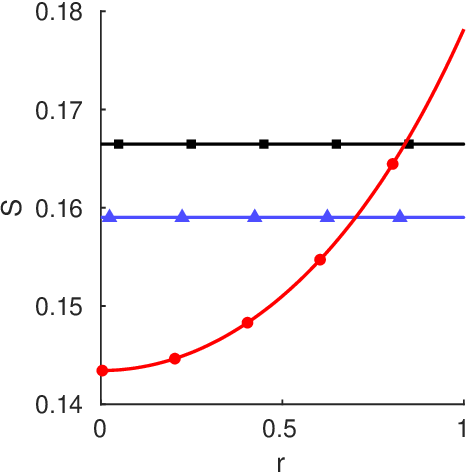}
\hspace*{0.15cm}
\raisebox{1.65cm}{\includegraphics[width=0.12\textwidth]{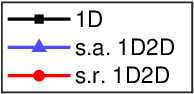}}
\hspace*{0.84cm}
\caption{Temperature $T$, crystallinity $x$ and orientational tensor component $S$ at outlet for all three models.}
\label{fig:nylonTestCaseOutletSolutions}
\end{figure}

The simulation results are visualized in Fig.~\ref{fig:nylonTestCaseAxialSolutions} and Fig.~\ref{fig:nylonTestCaseOutletSolutions}.
Across all models, the fiber cools down along the spinline from 550\,K at the inlet to about 300\,K and almost reaches the quench air temperature of 297.15\,K. Crystallization occurs primarily in the first half of the spinline and the crystallinity reaches a plateau between 0.065 and 0.08 for all models. The low value can be attributed to the low take-up velocity and the resulting low stress.
Regarding the cross-sectionally averaged fiber quantities, the outcomes of the three models are qualitatively and quantitatively similar, cf.~Fig.~\ref{fig:nylonTestCaseAxialSolutions}. The results of the 1D model, however, show slight but noticeable deviations from the results of the 1D2D models: at the ending of the spinline the temperature and crystallinity of the 1D model are lower and the semi-crystalline orientation of the 1D model is higher than the respectively averaged quantities of the 1D2D models. This comes from the impact of the radially resolved temperature on $x$ and $S$ in case of the 1D2D models.
The 1D2D models show similar temperatures at fiber center and surface along the spinline. Moreover, their radial temperature profiles at the take-up point coincide. Due to the outer cooling, the crystallization starts at the fiber surface. The fiber core cools down more slowly and thus begins to crystallize later. Hence, at the outlet a higher degree of crystallization is obtained at the fiber surface than in the core (sheath-core structure), cf.~Fig.~\ref{fig:nylonTestCaseOutletSolutions}. 
Despite of the low flow-enhanced crystallization effects, differences between the two 1D2D models can be observed: 
the stress-averaged 1D2D model predicts smaller variation in the crystallinity between fiber surface and center than the stress-resolved 1D2D model due to the averaged semi-crystalline orientation.
In terms of effort, the computation of all three models has an initialization step of about 210 seconds.
The radial coupling step that is additionally required for the 1D2D models takes about 90 seconds for the stress-averaged 1D2D model and about 220 seconds for the stress-resolved 1D2D model.

\begin{figure}[t]
\centering
\includegraphics[width=0.24\textwidth]{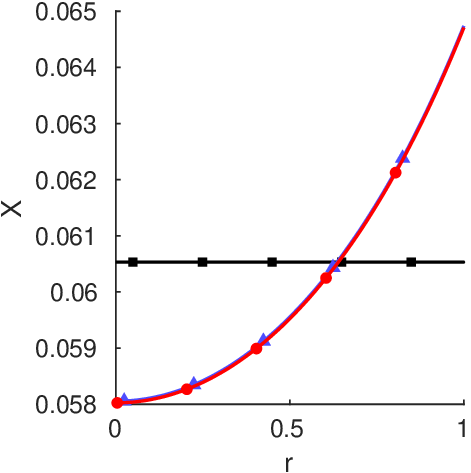}
\includegraphics[width=0.24\textwidth]{pictures/realTestCase/standard_FEC/outletSolution_crystallinity.eps}
\includegraphics[width=0.24\textwidth]{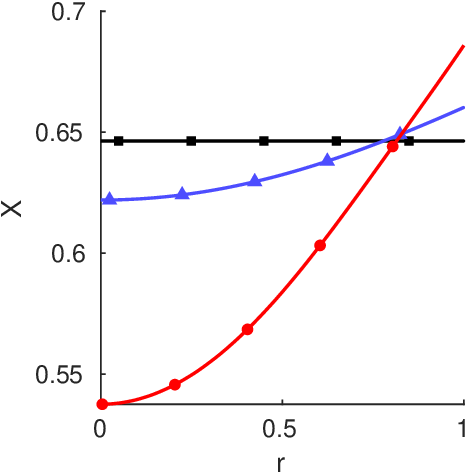}
\hspace*{0.1cm}
\raisebox{1.2cm}{\includegraphics[width=0.12\textwidth]{pictures/realTestCase/standard_FEC/legendRadial.eps}}
\caption{Crystallinity $x$ at outlet for low, standard and high FEC parameter $\xi\in7.2\cdot \{10^{-5},10^{-2},10^{-1}\}$ (from left to right). Note that the scales differ.}
\label{fig:nylonTestCaseCompareFecOutletSolutions}
\end{figure}

To analyze the impact of flow-enhanced crystallization on the spinning process, we perform a parameter study with three different values for the FEC parameter $\xi$. The higher $\xi$, the higher is the cross-sectionally averaged crystallinity $x$. The effect on the radial profile of $x$ at the outlet can be seen in Fig.~\ref{fig:nylonTestCaseCompareFecOutletSolutions}. The outcomes of the 1D2D models coincide for small $\xi$, as the radial resolution of the stress components has no significant influence on the crystallization. If the flow-enhanced crystallization is strong in case of either a large $\xi$ or generally high stresses due to higher take-up velocities, the results of the 1D2D models differ. The stress-resolved 1D2D model predicts a sheath-core structure with distinctive radial differences, whereas the differences predicted in the stress-averaged 1D2D are significantly smaller. Here, the information about radial stress differences is lost due to the additional cross-sectional averaging of the stress components. 

It should be noted that the assumption on the closure (boundary) conditions \eqref{eq:bcConstDoufas} for the stress-resolved 1D2D model could neither be verified nor falsified. Summing up, our novel stress-averaged 1D2D model provides equivalent results to the stress-resolved 1D2D model in cases of low flow-enhanced crystallization. It is faster and does not require any additional assumptions on the boundary conditions. It is therefore an alternative to be considered, especially in such scenarios. The 1D model is preferred for high thermal conductivities, where radial differences are negligibly small.

\section{Conclusion} \label{sec:conclusion}
The novel stress-averaged one-two-dimensional model presented in this paper is a conclusive extension of the model hierarchy for viscoelastic two-phase fiber melt spinning. It provides fast and reliable results, especially in the regime of low flow-enhanced crystallization, without requiring any further assumptions on the inlet profiles. While the presented model comparison between the dimensionally reduced models and the 3D reference indicates that the inlet profile assumption in \cite{doufas_two-dimensional_2001} seems reasonable for the considered amorphous test case, it could not be verified or falsified in general for the two-phase model. In terms of efficiency, all dimensionally reduced models achieve a significant acceleration of the calculations up to three orders of magnitude compared to the 3D model.

\section*{Acknowledgment}
The authors thank Dr.\ Sebastian Blauth at Fraunhofer ITWM for the invaluable insights and discussions on numerical techniques for partial differential equations and the finite element software FEniCS.

\appendix
\renewcommand{\theequation}{\Alph{section}.\arabic{equation}}
\renewcommand{\thetable}{\Alph{section}.\arabic{table}}
\renewcommand{\thefigure}{\Alph{section}.\arabic{figure}}
\setcounter{equation}{0} \setcounter{figure}{0} \setcounter{table}{0}

\section{Model derivation} \label{appendix:derivation}
The appendix provides details to the derivation and non-dimensionalization of the stress-resolved 1D2D model (System~\ref{system_fullradial}). We adapt here the derivation of a 1D2D fiber model for dry spinning from \cite{wieland_efficient_2019}, for the underlying asymptotic concepts we refer to \cite{panda_asymptotic_2008}.
For the dimensional reduction of the 3D free boundary value problem (System~\ref{system_3d}) it is essential to formulate the problem in appropriate fiber coordinates. Let $\boldsymbol{\breve{r}}: \mathcal{Q}_f \subset \mathbb{R}^3\rightarrow  \mathcal{Q}\in \mathbb{E}^3$ be a bijective mapping that maps fiber coordinates $\vece{x}$ onto a spatial point $\vece{r}$ in the Euclidian space, whereas its inverse maps any point onto its coordinates. Consequently, the scalar, vector and tensor fields of System~\ref{system_3d} can be defined in spatial points or fiber coordinates. To keep the terminology simple, we apply the same notations for the fields independent of their definition domain. 

\subsection{Dimensional reduction} Consider a uniaxial, radially symmetric spinning setup in the three-dimensional Euclidian space $\mathbb{E}^3$. We introduce a fixed orthonormal basis $\lbrace\vece{a}_1,\vece{a}_2,\vece{a}_z\rbrace \in \mathbb{E}^3$, where $\vece{a}_z$ points from the inlet (nozzle) in the direction of the outlet, implying the gravity to hold $\vece{g} = \rho g \vece{a}_z$.

\begin{assumption}[Fiber geometry] \label{assump:uniaxial_radialsymmetric} 
The fiber domain $\mathcal{Q}_f $ is given by the fiber length $L$ and the smooth radius function $R:[0,L]\rightarrow \mathbb{R}^+$ in such a way that
$$ \mathcal{Q}_f= \{ \vece x=(x_1,x_2,s)\in \mathbb{R}^3 \, | \, (x_1,x_2)\in \mathcal{A}(s), \, s \in [0,L] \} $$
with cross-sections $\mathcal{A}(s) = \{ (x_1,x_2)\in \mathbb{R}^2 \, | \, x_1 = r \cos \phi, x_2 = r \sin \phi, \, r \in [0,R(s)], \phi \in [0, 2\pi) \}$.
\end{assumption}
Then, the lateral surface of the fiber can be parametrized by the bijective function $\vece{\xi}(\phi,s)=(R(s) \cos \phi, R(s) \sin \phi, s)$ for $(\phi,s)\in [0,2\pi)\times [0,L]$, and the outer normal is given by
\begin{align*}
\vece{n}(\phi,s) = \frac{\partial_\phi \vece{\xi} \times \partial_s \vece{\xi}}{\norm{ \partial_\phi \vece{\xi} \times \partial_s \vece{\xi}}} (\phi,s) = \frac{1}{\sqrt{1+(\partial_s R(s))^2}} (\cos \phi, \sin \phi, -\partial_s R(s)).
\end{align*}
For any differentiable and integrable scalar-, vector- or tensor-valued function $f$ in $\mathcal{Q}_f$ we introduce the notation
\begin{align*}
\langle f \rangle_{\mathcal{A}(s) }&= \int_{\mathcal{A}(s)} f(x_1,x_2,s) \,\mathrm{d}x_1 \mathrm{d}x_2, \qquad \langle f \rangle_{\partial \mathcal{A}(s)} = \int_{\partial \mathcal{A}(s)} \frac{f}{\sqrt{n_1^2 + n_2^2}}\, \mathrm{d}l
\end{align*}
according to \cite{panda_asymptotic_2008}. The Gauss theorem yields the averaging rule
\begin{align}
\avg{\nabla \cdot f} = \partial_s \avg{f^T \cdot \vece{e}_3} + \avgBound{f^T \cdot \vece{n}}.
\end{align}

\subsubsection*{Averaging strategy and assumptions}
The 1D2D modeling pursues the idea of considering cross-sectionally averaged velocity information and radially resolved temperature information. This procedure requires an ansatz for certain variables to complete the cross-sectionally averaged momentum balance and to maintain a radially consistent energy balance.
\begin{assumption} \label{assump:velocity_pressure}
We consider:
\begin{enumerate}
\item[a)] the axial velocity only depends on the axial position,  $v_z(\vece{x}) = u(s)$;
\item[b)] the density varies only in axial direction;
\item[c)] the pressure equals the radial total extra stresses,  $p = \tau_{rr}$, cf.~\cite{yarin_fundamentals_2014};
\item[d)] the diffusive heat transfer in axial direction can be neglected.
\end{enumerate}
\end{assumption}
\begin{remark} \label{remark:densityAverage}
In case of a temperature and crystallinity-dependent density $\mathbf{\rho}$ in System~\ref{system_3d} we consider
$$\bar \rho(s)=\rho(\avgUnit{T}(s),\avgUnit{x}(s)), \qquad \avgUnit{T}=\avg{T}/A, \qquad \avgUnit{x}=\avg{x}/A$$
with the cross-sectionally averaged temperature and crystallinity in accordance to Assumption \ref{assump:velocity_pressure} b).
\end{remark}
The cross-sectionally averaged mass balance results in a constant mass flux $Q$,
\begin{align*}
\partial_s \avg{\rho v_z} &= 0, \qquad \rightarrow \qquad Q=\bar \rho u \, A  = const, \quad A= \avg{1}.
\end{align*}
Averaging the momentum balances and applying the lateral surface conditions, we obtain 
\begin{align*}
\partial_s \avg{\rho v_z^2} &= \partial_s \avg{\left(\tensor{\Sigma} \cdot \vece{e}_3\right) \cdot \vece{e}_3} + \avgBound{\vece{f}_\mathrm{air} \cdot \vece{e}_3}+  \avgBound{\vece{f}_\mathrm{st} \cdot \vece{e}_3} + \avg{\rho g}.
\end{align*}
The stress tensor $\tensor{\Sigma}=-p\tensor{I}+\tensor{\tau}$ consists of pressure $p$ and the total extra stress $\tensor{\tau}$ resulting from the amorphous and semi-crystalline extra stresses, $\tensor{\tau}=\tensor{\tau}_\mathrm{am} + \tensor{\tau}_\mathrm{sc}$. The averaged axial force due to surface tension can be expressed in terms of the surface tension coefficient $\gamma$ and the radius function $R$, \cite{marheineke_asymptotic_2009}. Using
\begin{align*}
\avg{\left(\tensor{\Sigma} \cdot \vece{e}_3\right) \cdot \vece{e}_3} = \avg{-p + \tau_{zz}}
 \qquad f_\mathrm{st} = \avgBound{\vece{f}_\mathrm{st} \cdot \vece{e}_3}=\pi\gamma \partial_s R, \qquad  f_\mathrm{air} = \avgBound{\vece{f}_\mathrm{air} \cdot \vece{e}_3},
\end{align*}
the averaged momentum balance becomes
\begin{align*}
Q \partial_s u &= \partial_s \avg{\tau_{zz}-\tau_{rr}} + f_\mathrm{air} + f_\mathrm{st} + \bar \rho g A.
\end{align*}

\subsubsection*{Radial resolution}
The radial symmetry of the fiber allows the restriction to a two-dimensional cutting plane spanned by $\vece{a}_z$ and the orthonormal radial vector $\vece{a}_r$, implying the fiber coordinates
$$Q_{cut} = \{(s,r) \in \mathbb{R}^2\,|\, r \in [0, R(s)], \, s \in [0, L]\}.$$ 
Using Assumption~\ref{assump:velocity_pressure} a) and b) we can conclude from the original three-dimensional mass balance in cylinder coordinates and its cross-sectionally averaged version that the radial velocity $v_r$ is linear in the radial component $r$ and can be expressed in terms of the axial velocity as
\begin{align*}
v_r(s,r) = r \frac{\partial_s R(s)}{R(s)} u(s).
\end{align*}
Considering the energy balance in cylinder coordinates on $\mathcal{Q}_{cut}$, we hence obtain
\begin{align*}
 c_p(T,x) \bar\rho u \left(\partial_s T + r \frac{\partial_s R}{R} \partial_r T \right) - \frac{1}{r} \partial_r \left(r C(T,x) \partial_r T \right) = \Phi_\infty \Delta H_f(T,x) \, \bar\rho u \left(\partial_s x + r \frac{\partial_s R}{R} \partial_r x \right).
\end{align*}
For computational reasons, we linearize the advection-diffusion equation around the cross-sectionally averaged temperature and crystallinity, yielding
\begin{align*}
 c_p(\avgUnit{T},\avgUnit{x}) \bar\rho u \left(\partial_s T + r \frac{\partial_s R}{R} \partial_r T \right) &- C(\avgUnit{T},\avgUnit{x})\frac{1}{r} \partial_r \left(r\partial_r T \right) \\
&= \Phi_\infty \Delta H_f(\avgUnit{T},\avgUnit{x}) \bar\rho u \left(\partial_s x + r \frac{\partial_s R}{R} \partial_r x \right),
\end{align*}
with $\avgUnit{f} = \avg{f}/A$.

\subsubsection*{Crystallization and microstructural equations}
In the stress-resolved 1D2D fiber model the evolution equation for the crystallization, the stored free energy and the microstructural equations are considered on $\mathcal{Q}_{cut}$. Under the assumptions, we have for $x$
\begin{align*}
u  \left(\partial_s x + r \frac{\partial_s R}{R} \partial_r x \right) = K (1-x).
\end{align*}
The assumption of an uniaxial, radially symmetric fiber yields a diagonal form of the stress tensor $\tensor{\Sigma}$, the conformation tensors $\tensor{c}$ and the orientation tensor $\tensor{S}$.
Furthermore, it holds that $c_{rr} = c_{\phi\phi}$ and $S_{rr}=S_{\phi\phi}$.
Together with the vanishing trace of $\tensor{S}$ the microstructural equations result in differential equations for the tensor components $c_{zz}$, $c_{rr}$ and $S_{zz}$,
\begin{align*}
&\lambda_\mathrm{am} \left( u \partial_s c_{zz} + u \, r \frac{\partial_s R}{R} \partial_r c_{zz} - 2c_{zz}\partial_s u \right) 		= - \bigg((1-\alpha) + \frac{ \alpha}{\zeta} \frac{c_{zz}}{1-x} \bigg) \bigg(c_{zz} - \zeta (1-x) \bigg), \\
&\lambda_\mathrm{am} \left( u \partial_s c_{rr} + u \, r \frac{\partial_s R}{R} \partial_r c_{rr} - 2 c_{rr} u \frac{\partial_s R}{R} \right)		= - \bigg((1-\alpha) + \frac{ \alpha}{\zeta} \frac{c_{rr}}{1-x} \bigg) \bigg(c_{rr} - \zeta (1-x) \bigg), \\
&\lambda_\mathrm{sc} \left( u \partial_s S_{zz} + u \, r \frac{\partial_s R}{R} \partial_r S_{zz} - 2 S_{zz} \partial_s u \right)		=  - \sigma S_{zz} + 2 \lambda_\mathrm{sc} \left( \Big( \frac{1}{3} - \closureFzU(S_{zz}) \Big) \partial_s u - \closureFzRho(S_{zz}) u \frac{\partial_s \rho}{\rho} \right).
\end{align*}
The stored free energy $a$ is then given by
\begin{align*}
u \left( \partial_s a + r \frac{\partial_s R}{R} \partial_r a \right) &= - \frac{1}{\lambda_\mathrm{am}} a + \frac{G}{\zeta} \frac{1}{1-x} \left( c_{zz} \partial_s u + 2 c_{rr} \frac{\partial_s R}{R} u \right).
\end{align*}

\subsection{Non-dimensionalization and scaling}
For the numerical treatment it is convenient to consider the fiber models in non-dimensional form and unit domains.
Therefore, we introduce for each dimensional quantity $y$ a dimensionless one $y^*$ as $y^*(s^*) = y(s_0 s^*)/y_0$ or $y^*(s^*,r^*) = y(s_0 s^*, R_0 r^*)/y_0$, respectively. The reference values and the resulting dimensionless numbers are listed in Table~\ref{tab:nondim}. Consequently, $s^* \in [0,1]$ and $Q^* = 1$ hold.
After non-dimensionalization, the 2D equations are formulated on the radius-dependent domain $[0,1] \times [0,R^*(s^*)]$. Applying the transformation $\hat{y}(s^*,\hat{r}) = y^*(s^*,R^*(s^*)\hat{r})$, the 2D equations can be considered on the unit square $[0,1]^2$. Note that the transformation onto the unit square eliminates all terms containing the radius function and its derivative in the 2D equations. As a consequence, the microstructural equations depend only parametrically on the radial component in the stress-resolved 1D2D fiber model due the radially resolved temperature.

The cross-sectionally averaged momentum balance contains a second order derivative of the axial velocity $u$ due to the stress tensor components $\tau_{zz}$ and $\tau_{rr}$. We introduce $\omega = \partial_s u$ as a new variable to get an explicit system of first order. Furthermore, we introduce the scaled unknown $S = S_{zz}/\delta$ for asymptotic reasons \cite{ettmuller_asymptotic_2021}. 

\section{Closure Approximation for semi-crystalline phase}\label{appendix:closure}
The closure approximation tensor $\tensor{U}$ and the closure approximation terms $\closureFzU(S_{zz})$, $\closureFrU(S_{zz})$, $\closureFzRho(S_{zz})$ and $\closureFrRho(S_{zz})$ for the semi-crystalline phase in Section~\ref{sec:3D} and Section~\ref{sec:reducedModels} result from microstructural considerations, see \cite{doufas_simulation_2000-1} for details.
The tensor $\tensor{U}$ is given by
\begin{align*}
\tensor{U} &= (1-w) \left(\frac{1}{15} \left( \nabla\vece{v} + (\nabla\vece{v})^T\right) + \frac{1}{7} \left( \langle \nabla \vece{v} , \tensor{S} \rangle_F \tensor{I} + \tensor{S} \cdot \left(\nabla\vece{v} + (\nabla\vece{v})^T\right) + \left(\nabla\vece{v} + (\nabla\vece{v})^T\right) \cdot \tensor{S} \right) \right) \\
&\,\, \quad + w \langle \nabla \vece{v} , \tensor{S} \rangle_F \left(\tensor{S}+\frac{1}{3}\tensor{I}\right), \\
w &= 1-27\det(\tensor{S}+\frac{1}{3}\tensor{I}).
\end{align*}
The operator $\langle \cdot , \cdot \rangle_F$ denotes the Frobenius inner product defined as $\langle A, B \rangle_F = \sum_{i=1}^m \sum_{j=1}^n A_{ij} B_{ij}$ for arbitrary matrices $A,B \in \mathbb{R}^{m\times n}$ with $n,m \in \mathbb{N}$.
For the dimensionally reduced models, the tensor components $U_{zz}$ and $U_{rr}$ are
\begin{align*}
U_{zz} = \closureFzU(S_{zz}) \partial_s u + \closureFzRho(S_{zz}) u \frac{\partial_s \rho}{\rho}, \\
U_{rr} = \closureFrU(S_{zz}) \partial_s u + \closureFrRho(S_{zz}) u \frac{\partial_s \rho}{\rho},
\end{align*}
where the closure approximation terms can be expressed as polynomials in the orientational tensor component $S_{zz}$, i.e.,
\begin{align*}
\closureFzU(S_{zz})		&= -\frac{81}{8}S_{zz}^5 + \frac{675}{56}S_{zz}^4 - \frac{36}{35}S_{zz}^3 - \frac{9}{10}S_{zz}^2 + \frac{11}{14}S_{zz} + \frac{2}{15}, \\
\closureFrU(S_{zz})		&= \phantom{-}\frac{81}{16}S_{zz}^5 - \frac{675}{112}S_{zz}^4 + \frac{18}{35}S_{zz}^3 + \frac{9}{20}S_{zz}^2 + \frac{5}{14}S_{zz} - \frac{1}{15}, \\
\closureFzRho(S_{zz})		&= -\frac{27}{8}S_{zz}^5 + \frac{153}{56}S_{zz}^4 + \frac{9}{14}S_{zz}^3 + \frac{1}{14}S_{zz}, \\
\closureFrRho(S_{zz})		&= \phantom{-}\frac{27}{16}S_{zz}^5 - \frac{153}{112}S_{zz}^4 - \frac{27}{35}S_{zz}^3 + \frac{9}{20}S_{zz}^2 + \frac{3}{14}S_{zz} - \frac{1}{15}.
\end{align*}

\section{Melt spinning of Nylon-66: Closing models and parameters}\label{appendix:nylon}
Melt spinning of Nylon-66 is considered in Section \ref{sec:nylonCase}. The closing models / functions and parameters used for the two-phase fiber models are briefly stated in the following. The process parameters of the specific test case are listed in Table~\ref{tab:nylon_process_parameters}.

\subsection{Material properties}
The models for the fiber density $\rho$, dynamic viscosity $\mu$, specific heat capacity $c_\mathrm{p}$, thermal conductivity $C$ and specific latent heat of crystallization $\Delta H_\mathrm{f}$ of a Nylon-66 fiber are taken from \cite{doufas_two-dimensional_2001, shrikhande_modified_2006}. The fiber density $\rho$ and the thermal conductivity $C$ are considered to be constant;
\begin{alignat*}{2}
\mu(T)	&= \mu_{\mathrm{ref}} \exp\left(\frac{E_\mathrm{A}}{E_1} \frac{T_1 + T_2 - T}{T}\right), \\
c_\mathrm{p}(T,x)		&= c_\mathrm{p}^{\mathrm{cr}}(T) \,x \Phi_{\infty} + c_\mathrm{p}^{\mathrm{am}}(T) \,\left(1-x \Phi_{\infty}\right), \\
& \,\, \quad c_\mathrm{p}^{\mathrm{cr}}(T)	\,\,= c_{\mathrm{s}1} + c_{\mathrm{s}2} (T-T_1), \\
& \,\, \quad c_\mathrm{p}^{\mathrm{am}}(T)	= c_{\mathrm{l}1} + c_{\mathrm{l}2} (T-T_1), \\
\Delta H_\mathrm{f}(T)	&= \Delta H_\mathrm{ref} + (c_{\mathrm{l}1}-c_{\mathrm{s}1})(T-T_1) + (c_{\mathrm{l}2}-c_{\mathrm{s}2}) \frac{(T-T_1)^2}{2}.
\end{alignat*}
For the values of referential viscosity $\mu_{\mathrm{ref}}$ [Pa s], activation energy $E_\mathrm{A}$ [J/mol], referential heat of crystallization $\Delta H_{\mathrm{ref}}$ [J/kg] and ultimate degree of crystallization $\Phi_{\infty}$ see Table~\ref{tab:nylon_physical_parameters}; the remaining parameters are 
\begin{alignat*}{3}
&  T_1 = 273.15 \;\tn{K}, 		&& T_2 = 280 \; \tn{K}, && \quad E_1 = 4599.05 \; \tn{J/mol},\\
& c_{\mathrm{s}1} = 1.255 \cdot 10^{3}\; \tn{J/(kg K)}, 		&& c_{\mathrm{s}2} = 8.368 \; \tn{J/(kg K}^2), \\
& c_{\mathrm{l}1} = 2.092 \cdot 10^{3} \; \tn{J/(kg K}), \qquad && c_{\mathrm{l}2} = 1.946 \;  \tn{J/(kg K}^2).
\end{alignat*}
The relaxation time $\lambda$ is described with a constant shear modulus $G$ as
 \begin{align*}
 \lambda(T) = \frac{\mu(T)}{G}, \qquad \quad
\deriv{\lambda} = - \lambda \frac{E_\mathrm{A}}{E_1} \frac{T_1+T_2}{T^2} \deriv{T}.
\end{align*}
The chosen values for the other model parameters that are associated to the two phases and the crystallization are listed in Table~\ref{tab:shrikhande_model_parameters}.

\begin{table}[tb]
\centering
\begin{tabular}{|l l l l|}
\hline
\multicolumn{4}{|l|}{\textbf{Physical and rheological parameters}}						\\
Description								& Symbol			& Value		& Unit			\\
\hline
Density									& $\rho$			& 1106 		& kg/$\tn{m}^3$ \\
Thermal conductivity					& $C$				& 0.209		& W/(m K)		\\
Referential viscosity at 280$^\circ$C	& $\mu_\text{ref}$		& 228.1528	& Pa s			\\
Activation energy						& $E_\text{A}$				& $5.6484 \cdot 10^4$	& J/mol \\
Referential heat of crystallization				& $\Delta H_\text{ref}$	& $2.0920 \cdot 10^5$ 	& J/kg  \\
Ultimate degree of crystallization	& $\Phi_{\infty}$	& 0.3		& -				\\
Maximum crystallization rate			& $K_\text{max}$			& 1.64		& 1/s			\\
Temperature of maximum crystallization rate & $T_\text{max}$		& 423.15	& K				\\
Temperature half-width in crystallization rate					& $\Delta T$				& 80		& K				\\
Shear modulus						& $G$		& $1.1 \cdot 10^5$	& Pa	\\ 
Surface tension							& $\gamma$	& 0.036		& N/m			\\
\hline
\end{tabular}
\caption{Physical and rheological parameters for Nylon-66 melt, \cite{doufas_two-dimensional_2001, shrikhande_modified_2006}. Note that no value for $\zeta=N_0l^2/3$ [m$^2$] is given in the literature. Since it does not play a role in the dimensionless model variants, we also do not specify it. }
\label{tab:nylon_physical_parameters}
\end{table}

\begin{table}[tb]
\centering
\begin{tabular}{|l l l |}
\hline
\multicolumn{3}{|l|}{\textbf{Model parameters for two-phase flow}}						\\
Description								& Symbol			& Value			\\
\hline
Giesekus mobility parameter								& $\alpha$			& 1.0 \\
Anisotropic drag coefficient	& $\sigma$		& 1.0				\\
Parameter for semi-crystalline relaxation time & $F$				& 20	 \\
Parameter for semi-crystalline relaxation time & $\delta$			& 0.02	\\
Parameter for flow-enhanced crystallization				& 	& 	\\
\qquad \quad low &$\xi$ & $7.2 \cdot 10^{-5}$\\
\qquad \quad standard \quad &$\xi$ & $7.2 \cdot 10^{-2}$\\
\qquad \quad high \quad &$\xi$ & $7.2 \cdot 10^{-1}$\\
\hline
\end{tabular}
\caption{Model parameters associated to the two phases and the crystallization for melt spinning of Nylon-66, \cite{doufas_two-dimensional_2001, shrikhande_modified_2006}. }
\label{tab:shrikhande_model_parameters}
\end{table}

\begin{table}[tb]
\centering
\begin{tabular}{|l l l l|}
\hline
\multicolumn{4}{|l|}{\textbf{Process parameters}}											\\
Description								& Symbol			& Value		& Unit			\\
\hline
Fiber length							& $L$				& 1.5		& m				\\
Nozzle diameter							& $D_\text{in}$			& $3.81 \cdot 10^{-4}$ & m	\\
Temperature at inlet					& $T_\text{in}$			& 550.15	& K				\\
Velocity at inlet						& $u_\text{in}$			& 0.1877	& m/s			\\
Take-up velocity at outlet						& $v_\text{out}$			& $16.7\overline{6}$		& m/s			\\
\hline
Air temperature							& $T_\mathrm{air}$			& 297.15	& K \\
Air velocity			                & $v_\mathrm{air}$			& 0.4		& m/s \\
Air density								& $\rho_\mathrm{air}$			& 1.0		& kg/$\tn{m}^3$\\
Air specific heat capacity				& $c_{\mathrm{p},\mathrm{air}}$			& $1.0 \cdot 10^{3}$		& J/(kg K) \\
Air thermal conductivity				& $k_\mathrm{air}$			& $3.1\cdot 10^{-2}$		& W/(m K)\\
Air kinematic viscosity					& $\nu_\mathrm{air}$			& $2.0 \cdot 10^{-5}$		& $\tn{m}^2$/s \\
\hline
\end{tabular}
\caption{Process parameters for melt spinning of Nylon-66, \cite{doufas_two-dimensional_2001}.}
\label{tab:nylon_process_parameters}
\end{table}

\subsection{Aerodynamic drag and heat transfer}
The aerodynamic drag forces $f_\tn{air}$ on the fiber dynamics are described by the air drag model of \cite{marheineke_dragmodel_2011}. The model for the heat transfer coefficient $\alpha_T$ is taken from \cite{wieland_efficient_2019}. The required air quantities are considered to be constant and are given in Table~\ref{tab:nylon_process_parameters}.

\section{Numerical treatment}\label{appendix:numerics}
\subsection{Solver for three-dimensional model}
To numerically solve the three-dimensional fiber model given by System~\ref{system_3d}, we use the open-source finite element software FEniCS, which basically takes a weak formulation of System~\ref{system_3d}, cf.~\cite{alnaes_fenics_2015,logg_fenics_2012}. We use artificial diffusion to stabilize the numerical solver by modifying the momentum balance and the constitutive equation for the conformation tensor $\tensor{c}$
\begin{align*}
\nabla \cdot (\rho \vece{v} \otimes \vece{v}) &= \nabla \cdot \tensor{\Sigma}^T + \vece{g} + \varepsilon_1 \Delta \vece{v}, \\
\lambda_\mathrm{am} \; \overset{\triangledown}{\tensor{c}} &= - \Bigg( (1-\alpha)\tensor{I} + \frac{\alpha}{\zeta} \frac{1}{1-x} \tensor{c}\Bigg) \Bigg( \tensor{c} - \zeta (1-x) \tensor{I} \Bigg) + \varepsilon_2 \Delta \tensor{c},
\end{align*}
with Laplacian operator $\Delta$ and stabilization parameters $\varepsilon_1 = 10^{-2}$ and $\varepsilon_2 = 10^{-12}$.
We transform the weak formulation to the fixed domain $\tilde{\mathcal{Q}}_{cut} = [0,L] \times [0,1]$ such that no re-meshing is needed. The fiber radius $R$ can thereby be calculated from the constant mass flow $Q$. The solution algorithm then iteratively solves the combined equations for mass, momentum and conformation tensor with the equation for the fiber radius and the energy balance.
The problem is solved on a triangular grid which is refined at the inlet $\Gamma_\tn{in}$ with approximately $1.1 \cdot 10^5$ triangular elements in total.

\subsection{Solvers for dimensionally reduced models}
The 1D2D fiber models are structured into three parts with respect to the type of differential equations.
Applying the method of lines approach to the equations \eqref{eqFull:onetwo_material} of the stress-resolved fiber model on a radial grid $0=r_1,\ldots,r_{N_r}=1$, $N_r \in \mathbb{N}$ yields $5N_r$ ordinary differential equations in $s$. Together with the cross-sectionally averaged equations \eqref{eqFull:onetwo_averaged} and the respective boundary conditions, we arrive at a boundary value problem of ordinary differential equations with $5N_r+2$ unknowns.
Analogously, we obtain a boundary value problem of ordinary differential equations with $N_r+6$ unknowns for the equations \eqref{eqInter:onetwo_averaged} and \eqref{eqInter:onetwo_material} of the stress-averaged fiber model.
The temperature equations \eqref{eqFull:onetwo_diffusion} and \eqref{eqInter:onetwo_diffusion} are partial differential advection-diffusion equations depending on $s$ and $r$.
The general solution algorithm is then given by Algorithm~\ref{algo}.
Note that the initialization step corresponds to solving the cross-sectionally averaged 1D model (System~\ref{system_averaged}). \\[-0.2cm]
\begin{algorithm} ~\newline\\[-0.8cm] \label{algo}
\begin{enumerate}[label=(\arabic*)]
\item \textit{Initialization step}: Compute the initial solution $\bs{y}^{(0)}$ solving the 1D model (System \ref{system_averaged}).
\item \textit{Coupling step}: For $i\geq 1$:
\begin{enumerate}
\item Solve equation \eqref{eqFull:onetwo_diffusion} or \eqref{eqInter:onetwo_diffusion} to obtain $T^{(i)}$.
\item Solve equations \eqref{eqFull:onetwo_averaged} together with \eqref{eqFull:onetwo_material} or \eqref{eqInter:onetwo_averaged} together with \eqref{eqInter:onetwo_material} to obtain in total $\bs{y}^{(i)}$.
\item STOP, if $\lvert\lvert \bs{y}^{(i-1)}-\bs{y}^{(i)} \rvert\rvert <tol$, $tol \in \mathbb{R}^+$.
\end{enumerate}
\end{enumerate}
\end{algorithm}
\subsubsection*{Initialization step}
The boundary value problem of  ordinary differential equations given by System~\ref{system_averaged}, we employ a con\-tin\-u\-a\-tion-collocation scheme implemented in MATLAB\footnote{For details on MATLAB, we refer to https://mathworks.com} which has been already used in \cite{ettmuller_asymptotic_2021}.
We embed the boundary value problem into a family of problems by introducing a continuation parameter vector $\bs{p} \in [0,1]^7$. The starting solution to the parameter vector $\bs{p}=\bs{0}$ is given by a stress-free, non-crystallizing fiber with constant temperature and velocity, i.e.\
\begin{alignat*}{7}
u &\equiv 1, \qquad\qquad	 & T &\equiv1, \qquad\qquad	& c_{zz} &\equiv1, \qquad\qquad & c_{rr} &\equiv1 \\
\omega &\equiv 0,  & S &\equiv0,				& x &\equiv 0, 	& a & \equiv 0.
\end{alignat*}
The continuation path is given by a four-staged solution strategy to incorporate the effects related to gravity, air drag, drawing, temperature, crystallization and model parameters step by step. For more details, we refer to \cite{ettmuller_asymptotic_2021}.
\subsubsection*{Coupling step}
In the coupling step, the advection-diffusion equations \eqref{eqFull:onetwo_diffusion} and \eqref{eqInter:onetwo_diffusion} are solved with FEniCS using linear Lagrange finite elements on a uniform triangular mesh with size $\Delta x = 10^{-3}$ in axial direction and size $\Delta r = 10^{-2}$ in radial direction.
Alternatively, an approach based on the analytical solution and the product integration method could be used, see \cite{ettmuller_productIntegration_2023}.
For the radial grid point $r_i$, $i=1,\ldots,N_r$, we use the nodes of the Gauß-Legendre quadrature with $N_r=20$. Interpolation between radial nodes of the advection-diffusion equation solver and the radial nodes of the method of lines technique is performed using piecewise cubic interpolation based on the modified Akima interpolation.
The boundary value problem of ordinary differential equations given by \eqref{eqFull:onetwo_averaged} and \eqref{eqFull:onetwo_material} is solved using the continuation-collocation scheme, analogously for \eqref{eqInter:onetwo_averaged} and \eqref{eqInter:onetwo_material}.
Adapting the continuation strategy of \cite{wieland_efficient_2019}, we replace the dynamic viscosity $\avgUnit{\mu(T)}^{(i)}$ in iteration $i$ by a linear combination,
\begin{equation*}
p_r\avgUnit{\mu(T)}^{(i)} + (1-p_r)\avgUnit{\mu(T)}^{(i-1)},
\end{equation*}
with radial continuation parameter $p_r \in [0,1]$. The dynamic viscosity in the first coupling iteration is initialized with $\avgUnit{\mu(T)}^{(0)} = \mu(\avgUnit{T})$.
The stopping criterion for the iteration in the coupling step is set to $tol = 10^{-6}$.

\bibliographystyle{siam}
\bibliography {bibmain}
\end{document}

%% file: Tikz/sketchRadialMeltSpinning.tikz
\tikzset{
    partial ellipse/.style args={#1:#2:#3}{
        insert path={+ (#1:#3) arc (#1:#2:#3)}
    }
}

\begin{tikzpicture}
\filldraw[fill=black!20!white, draw=black] (0,0) rectangle (2,-0.5) node[pos=0.5]{\small Spinneret};
\draw (3,0.3) node {\small Mass throughput};
\draw [-latex, line width=1.2] (3,0) -- (3,-0.5);
\filldraw[fill=black!20!white, draw=black] (4,0) rectangle (6,-0.5);

\draw (2,-0.5) to[out=180, in=100] (1.75,-1);
\draw (1.75,-1) to[out=-85, in=90] (2.9,-4.8);
\draw (2.9,-4.8) -- (2.9,-5.5);

\draw (4,-0.5) to[out=0, in=80] (4.25,-1);
\draw (4.25,-1) to[out=-95, in=90] (3.1,-4.8);
\draw (3.1,-4.8) -- (3.1,-5.5);

\draw (2.8,-5.58) -- (3.2,-5.52);
\draw (2.8,-5.62) -- (3.2,-5.56);
\draw (2.9,-5.65) -- (2.9,-7.5);
\draw (3.1,-5.65) -- (3.1,-7.5);

\draw[densely dotted] (3,-0.5) [partial ellipse=0:180:1.0 and 0.125];
\draw (3,-0.5) [partial ellipse=180:360:1.0 and 0.125];
\draw[align=left] (3.3,-0.5) node {\tiny $\Gamma_\mathrm{in}$};

\draw[densely dotted] (3,-3) [partial ellipse=0:180:0.57 and 0.165];
\draw (3,-3) [partial ellipse=180:360:0.57 and 0.165];
\draw[line width=0.1mm] (3,-3) -- (3.57,-3);
\draw[align=left] (3.18,-2.925) node {\tiny R};

\draw[densely dotted] (3,-6.72) [partial ellipse=0:180:0.1 and 0.05];
\draw (3,-6.72) [partial ellipse=180:360:0.1 and 0.05];
\draw (3,-6.72) -- (2.7,-6.8);
\draw[align=left] (2.6,-6.9) node {\tiny $\Gamma_\mathrm{out}$};

\draw[align=left] (1.8,-2) node {\tiny $\Gamma_\mathrm{fr}$};
\draw[align=left] (4.3,-2) node {\tiny $\Gamma_\mathrm{fr}$};

\draw [-latex, line width=0.5mm] (0,-3) -- (1.5,-3);
\draw [-latex, line width=0.5mm] (0,-3.75) -- (1.5,-3.75);
\draw [-latex, line width=0.5mm] (0,-4.5) -- (1.5,-4.5);
\draw (0.75, -2.5) node {\small Quench air};

\draw [stealth-stealth, line width=0.1mm] (4.75,-1) -- (4.75,-6.75);
\draw (4.5,-1) -- (5,-1) node[pos=2] {\small $s=0$};
\draw (4.5,-6.75) -- (5,-6.75) node[pos=2] {\small $s=L$};

\draw[thick,-latex] (3,-1) -- (3.5,-1) node[anchor=north] {\small r};
\draw[thick,-latex] (3,-1) -- (3,-1.5) node[anchor=east] {\small s};

\filldraw[fill=black!20!white, draw=black] (3.615,-7.5) circle (0.5);
\draw[align=left] (5.1,-7.5) node {\small Take-up \\ roll device};
\end{tikzpicture}

%% file: main.bbl
\begin{thebibliography}{10}

\bibitem{alnaes_fenics_2015}
{\sc M.~Aln{\ae}s, J.~Blechta, J.~Hake, A.~Johansson, B.~Kehlet, A.~Logg,
  C.~Richardson, J.~Ring, M.~E. Rognes, and G.~N. Wells}, {\em The {FEniCS}
  project version 1.5}, Archive of Numerical Software, 3 (2015).

\bibitem{doufas_simulation_2001}
{\sc A.~K. Doufas and A.~J. McHugh}, {\em Simulation of melt spinning including
  flow-induced crystallization. {Part} {III}. {Quantitative} comparisons with
  {PET} spinline data}, Journal of Rheology, 45 (2001), pp.~403--420.

\bibitem{doufas_two-dimensional_2001}
\leavevmode\vrule height 2pt depth -1.6pt width 23pt, {\em Two-dimensional
  simulation of melt spinning with a microstructural model for flow-induced
  crystallization}, Journal of Rheology, 45 (2001), pp.~855--879.

\bibitem{doufas_simulation_2000-1}
{\sc A.~K. Doufas, A.~J. McHugh, and C.~Miller}, {\em Simulation of melt
  spinning including flow-induced crystallization: {Part} {I}. {Model}
  development and predictions}, Journal of Non-Newtonian Fluid Mechanics, 92
  (2000), pp.~27--66.

\bibitem{doufas_simulation_2000}
{\sc A.~K. Doufas, A.~J. McHugh, C.~Miller, and A.~Immaneni}, {\em Simulation
  of melt spinning including flow-induced crystallization: {Part} {II}.
  {Quantitative} comparisons with industrial spinline data}, Journal of
  Non-Newtonian Fluid Mechanics, 92 (2000), pp.~81--103.

\bibitem{ettmuller_asymptotic_2021}
{\sc M.~Ettm{\"u}ller, W.~Arne, N.~Marheineke, and R.~Wegener}, {\em On
  flow-enhanced crystallization in fiber spinning: asymptotically justified
  boundary conditions for numerics of a stiff viscoelastic two-phase model},
  Journal of Non-Newtonian Fluid Mechanics, 296 (2021), p.~104636.

\bibitem{ettmuller_productIntegration_2023}
\leavevmode\vrule height 2pt depth -1.6pt width 23pt, {\em Product integration
  method for the simulation of radial effects in fiber melt spinning of
  semi-crystalline polymers}, PAMM, 22 (2023), p.~e202200210.

\bibitem{ettmueller_ecmi_2024}
\leavevmode\vrule height 2pt depth -1.6pt width 23pt, {\em Industrial melt
  spinning with two-way coupled airflow including viscoelasticity,
  crystallization and radial effects}, in Progress in Industrial Mathematics at
  ECMI 2023, Springer, 2024 (forthcoming).

\bibitem{henson_thin-filament_1998}
{\sc G.~M. Henson, D.~Cao, S.~E. Bechtel, and M.~G. Forest}, {\em A
  thin-filament melt spinning model with radial resolution of temperature and
  stress}, Journal of Rheology, 42 (1998), pp.~329--360.

\bibitem{hutchenson_radial_1984}
{\sc K.~W. Hutchenson, D.~D. Edie, and D.~M. Riggs}, {\em Radial temperature
  differences during the melt spinning of fibers}, Journal of applied polymer
  science, 29 (1984), pp.~3621--3640.

\bibitem{kohler_2d_2007}
{\sc W.~H. Kohler and A.~J. McHugh}, {\em {2D} {Modeling} of high-speed fiber
  spinning with flow-enhanced crystallization}, Journal of Rheology, 51 (2007),
  pp.~721--733.

\bibitem{logg_fenics_2012}
{\sc A.~Logg, K.-A. Mardal, and G.~Wells}, {\em Automated solution of
  differential equations by the finite element method: The FEniCS book},
  Springer Science \& Business Media, 2012.

\bibitem{marheineke_asymptotic_2009}
{\sc N.~Marheineke and R.~Wegener}, {\em Asymptotic model for the dynamics of
  curved viscous fibers with surface tension}, Journal of Fluid Mechanics, 622
  (2009), pp.~345--369.

\bibitem{marheineke_dragmodel_2011}
\leavevmode\vrule height 2pt depth -1.6pt width 23pt, {\em Modeling and
  application of a stochastic drag for fibers in turbulent flows},
  International Journal of Multiphase Flow, 37 (2011), pp.~136--148.

\bibitem{matsuo_studies_1976}
{\sc T.~Matsuo and S.~Kase}, {\em Studies on melt spinning. {VII}.
  {Temperature} profile within the filament}, Journal of Applied Polymer
  Science, 20 (1976), pp.~367--376.

\bibitem{ottone_numerical_2002}
{\sc M.~L. Ottone and J.~A. Deiber}, {\em A numerical method for the
  viscoelastic melt-spinning model with radial resolutions of temperature and
  stress fields}, Industrial \& engineering chemistry research, 41 (2002),
  pp.~6345--6353.

\bibitem{pan_radial_2019}
{\sc D.~Pan, H.~He, J.~Sun, M.~Cao, Z.~Qin, and L.~Chen}, {\em Radial
  crystallization difference of melt-spun polypropylene fiber along spinning
  line}, Journal of Applied Polymer Science, 136 (2019), p.~47175.

\bibitem{panda_asymptotic_2008}
{\sc S.~Panda, N.~Marheineke, and R.~Wegener}, {\em Systematic derivation of an
  asymptotic model for the dynamics of curved viscous fibers}, Mathematical
  Methods in the Applied Sciences, 31 (2008), pp.~1153--1173.

\bibitem{shrikhande_modified_2006}
{\sc P.~Shrikhande, W.~H. Kohler, and A.~J. McHugh}, {\em A modified model and
  algorithm for flow-enhanced crystallization---{Application} to fiber
  spinning}, Journal of Applied Polymer Science, 100 (2006), pp.~3240--3254.

\bibitem{sun_numerical_2000}
{\sc J.~Sun, S.~Subbiah, and J.~Marchal}, {\em Numerical analysis of
  nonisothermal viscoelastic melt spinning with ongoing crystallization},
  Journal of non-newtonian fluid mechanics, 93 (2000), pp.~133--151.

\bibitem{vassilatos_issues_1992}
{\sc G.~Vassilatos, E.~Schmelzer, and M.~Denn}, {\em Issues concerning the rate
  of heat transfer from a spinline}, International Polymer Processing, 7
  (1992), pp.~144--150.

\bibitem{wieland_efficient_2019}
{\sc M.~Wieland, W.~Arne, R.~Fe{\ss}ler, N.~Marheineke, and R.~Wegener}, {\em
  An efficient numerical framework for fiber spinning scenarios with
  evaporation effects in airflows}, Journal of Computational Physics, 384
  (2019), pp.~326--348.

\bibitem{yarin_fundamentals_2014}
{\sc A.~L. Yarin, B.~Pourdeyhimi, and S.~Ramakrishna}, {\em Fundamentals and
  applications of micro-and nanofibers}, Cambridge University Press, 2014.

\end{thebibliography}
